\documentclass[10pt,aps,prd,twocolumn,showpacs,superscriptaddress,nofootinbib,nobibnotes,longbibliography,floatfix]{revtex4-2}

\usepackage[utf8]{inputenc}
\usepackage[T1]{fontenc}
\usepackage{bm}
\usepackage{mathtools,amsmath,amssymb,amsfonts,mathrsfs,eucal,graphicx,tensor,csquotes,accents,commath,chngcntr,siunitx}
\usepackage[dvipsnames]{xcolor}
\usepackage[unicode]{hyperref}
\hypersetup{colorlinks=true, citecolor=MidnightBlue, linkcolor=MidnightBlue, urlcolor=MidnightBlue, linktocpage=true}
\usepackage[normalem]{ulem}
\usepackage{orcidlink}
\usepackage{graphicx}
\usepackage{float}

\newcommand{\orcid}[1]{\href{https://orcid.org/#1}{\includegraphics[width=10pt]{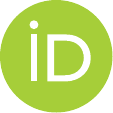}}}

\begin{document}

\author{Saulo Albuquerque \orcid{0000-0003-2911-9358}}
\affiliation{Departmento de Física, Universidade Federal da Paraíba, Caixa Postal 5008, João Pessoa 58059-900, Para\'iba, Brazil}
\affiliation{Theoretical Astrophysics, IAAT, University of Tübingen, D-72076 Tübingen, Germany}
\email{saulo.filho@academico.ufpb.br}

\author{Sebastian H. V\"olkel \orcid{0000-0002-9432-7690}}
\affiliation{Max Planck Institute for Gravitational Physics (Albert Einstein Institute), D-14476 Potsdam, Germany}
\email{sebastian.voelkel@aei.mpg.de}

\author{Kostas D. Kokkotas \orcid{0000-0001-6048-2919}}
\affiliation{Theoretical Astrophysics, IAAT, University of Tübingen, D-72076 Tübingen, Germany}

\date{\today}

\title{Inverse problem in energy-dependent potentials using semiclassical methods}

\begin{abstract}
Wave equations with energy-dependent potentials appear in many areas of physics, ranging from nuclear physics to black hole perturbation theory. In this work, we use the
semiclassical Wentzel-Kramers-Brillouin (WKB) method to first revisit the computation of bound states of potential wells and reflection/transmission coefficients in terms of the Bohr-Sommerfeld rule and the Gamow formula. We then discuss the inverse problem, in which the latter observables are used as a starting point to reconstruct the properties of the potentials. By extending known inversion techniques to energy-dependent potentials, we demonstrate that so-called width-equivalent or WKB-equivalent potentials are not isospectral anymore. Instead, we explicitly demonstrate that constructing quasi-isospectral potentials with the inverse techniques is still possible. Those reconstructed, energy-independent potentials share key properties with the width-equivalent potentials. We report that including energy-dependent terms allows for a rich phenomenology, particularly for the energy-independent equivalent potentials. 
\end{abstract}
\maketitle

\section{Introduction}\label{intro}

Energy-dependent potentials in wave equations play an important role in many different areas of physics. They appear naturally in nuclear physics~\cite{nuclear1,nuclear12,nuclear7,nuclear5,nuclear6,nuclear13,nuclear10,nuclear11,nuclear3,nuclear9,nuclear4,nuclear75,nuclear8,nuclear2}, when studying perturbations of black holes and neutron stars~\cite{Kokkotas:1999bd,Nollert:1999ji,Berti:2009kk,Konoplya:2011qq}, and in analog gravity~\cite{Barcelo:2005fc}. One popular approach to solving the wave equations is to use the Wentzel-Kramer-Brillouin (WKB) method. Among its most iconic tools are the classical Bohr-Sommerfeld rule~\cite{https://doi.org/10.1002/andp.19163561702} for the computation of bound states in potential wells, and the Gamow formula~\cite{Gamow:1928zz} for the computation of transmission and reflection coefficients. 

Due to their simplicity, it is known how they can be used to infer relevant information of the underlying potential in the inverse problem, i.e., when bound states or transmission coefficients can be provided~\cite{lieb2015studies,MR985100,1980AmJPh..48..432L,2006AmJPh..74..638G}. Such an inversion is also possible for one, three or four turning point potentials with quasistationary states~\cite{Volkel:2017kfj,Volkel:2018hwb,Volkel:2019gpq}. One of the key results is that the inversion is, in general, not unique. Instead, one can reconstruct a family of potentials with similar properties for their classical turning points. The universal property is that the separation of turning points must be unchanged. Thus, these potentials may also be called width equivalent potentials. In the literature, some authors have also coined such potentials as WKB-equivalent potentials; see Ref.~\cite{Bonatsos:1992qq}. 

To our knowledge, existing studies on the inverse problem using the Bohr-Sommerfeld rule or Gamow formula have only considered energy-independent potentials. However, many physical scenarios require one to work with energy-dependent potentials. Although WKB methods to solve the direct problem of bound states or transmission function are relatively straightforward to use, e.g., see Refs.~\cite{Kokkotas:1993ef,Konoplya:2003ii,Konoplya:2019hlu}, the energy dependence may introduce additional degeneracy for the inverse problem. 

In this work, we first revisit how the WKB method can be used for the direct and inverse problem for the energy-independent case. We then demonstrate how the standard methods for the inverse problem can be used to construct WKB-equivalent, energy-independent potentials from the bound states and transmission coefficients of energy-dependent potentials. As examples, we study extensions of the quadratic potential (harmonic oscillator), and the P\"oschl-Teller potential~\cite{Poschl:1933zz}. Using numerical methods, we also quantify how accurately these potentials can represent their energy-dependent pendants. One of our main findings is that WKB-equivalent potentials are not width-equivalent anymore. Another important finding is that those reconstructed, energy-independent, effective potentials capture some key properties of their associated energy-dependent ones, such as their asymptotic behavior and local behavior around their minimum or maximum. 

The rest of the paper is organized as follows. In Sec.~\ref{methods}, we review the semiclassical methods and our numerical scheme. Those are applied to two energy-dependent potentials in Sec.~\ref{app_results}. We discuss our findings in Sec.~\ref{discussion}, and our conclusions can be found in Sec.~\ref{conclusions}.

\section{Methods}\label{methods}

In the following, we first review some basics of the WKB method in Sec.~\ref{methods_wkb}, its application to potential wells in Sec.~\ref{methods_BS}, and potential barriers in Sec.~\ref{methods_GA}. Next, we discuss the important role played by the turning points within this framework in Sec.~\ref{methods_turning}, and finally, we introduce related numerical methods in Sec.~\ref{methods_NU}. 

\subsection{WKB method}\label{methods_wkb}

The WKB method, also known as semiclassical approximation, is a widely used approximation to study linear differential equations. A very common example in physics is the one-dimensional wave equation 
\begin{align}\label{wave_equation}
\frac{\text{d}^2}{\text{d}x^2}\psi(x) + Q(x,E)  \psi(x) = 0,
\end{align}
where
\begin{align}\label{wave_equation_potential}
Q(x,E)\equiv E-V(x). 
\end{align}

Here, $V(x)$ is an energy-independent potential, e.g., the harmonic oscillator. The WKB method is valid under several assumptions (see Refs.~\cite{bender1999advanced,2013waap.book.....K} as standard references for more details), and it breaks down close to classical turning points [defined by $Q(x,E)=0$, or equivalently, $E=V(x)$]. To construct solutions, one can connect exact, local solutions, e.g., described by the Airy functions, with the WKB solutions using asymptotic matching. One convenient application of that approach is to derive so-called quantization conditions to compute eigenvalues $E=E_n$ for given boundary conditions of a potential well or reflection/transmission coefficients through potential barriers. Although the WKB method is not exact, it can be an excellent approximation. Moreover, due to the integral equations, it is also possible to study the inverse problem, in which one is interested in reconstructing properties of the potential for given eigenvalues or transmission/reflection coefficients. We review two of the most commonly used applications in the subsequent sections.

\subsection{Classical Bohr-Sommerfeld rule}\label{methods_BS}

\begin{figure}
\includegraphics[width=1.0\linewidth]{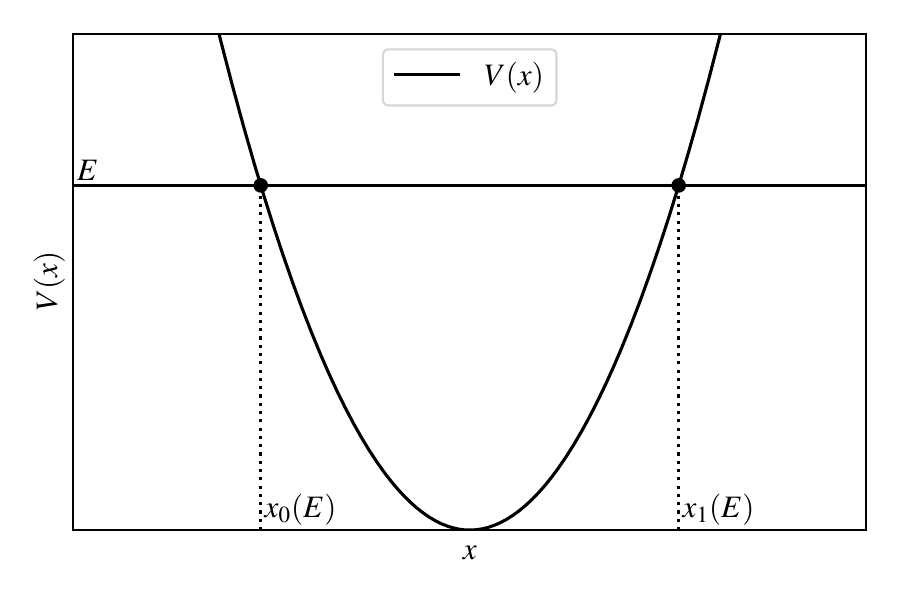}
\caption{
Potential well $V(x)$ with the pair of turning points $x_0(E), x_1(E)$ associated with the energy $E$ denoted as points at the intersection $E=V(x)$. \label{parabollicsimple}}
\end{figure}

The classical Bohr-Sommerfeld rule is given by
\begin{align}\label{cBS}
\int_{x_0}^{x_1} \sqrt{Q(x,E_{n})} \text{d}x = \pi \left(n+\frac{1}{2} \right),
\end{align}
where the classical turning points $x_0, x_1$ are defined via $Q(x,E_{n})=0$ and $n \in \mathrm{N}_0$. It can be derived by matching the WKB solutions of the classically allowed region [$E > V(x)$] and forbidden region [$E< V(x)$] with Kramer's matching relations close to the turning points. Thus, it applies to potential wells with two turning points, schematically shown in Fig.~\ref{parabollicsimple}. 

The Bohr-Sommerfeld rule can be inverted to study the inverse problem, in which one is provided with some spectrum $E_n$ and is interested in reconstructing the potential~\cite{lieb2015studies,MR985100}. Solutions to the inverse problem are, in general, not unique, and the problem itself may even be ill-posed. By inverting the Bohr-Sommerfeld rule, the nonuniqueness is imprinted in the observation that one can only reconstruct the width of the potential defined by the separation of turning points ,
\begin{align}\label{Excursion}
L_1(E) = x_1(E)-x_0(E) = \frac{\partial }{\partial E} I(E),
\end{align}
where $I(E)$ is the so-called inclusion
\begin{align}\label{Inclusion}
I(E) = 2 \int_{E_\text{min}}^{E}\frac{n(E^\prime)+1/2}{\sqrt{E-E^\prime}} \mathrm{d}E^\prime.
\end{align}
The minimum of the potential $E_\text{min}$ is not a part of the spectrum and has to be extrapolated from $n(E_\text{min}) = -1/2$. Infinitely many potentials admit the same spectrum $E_n$ and have the same width but can be ``tilted'' and ``shifted'' by the necessary freedom of providing one of the two turning-point functions.

\subsection{Gamow formula}\label{methods_GA}

The Gamow formula approximates the transmission $T(E)$ through a two turning point potential barrier $V(x)$ via
\begin{align}\label{Gamow}
T(E)  = \exp\left(2 \mathrm{i} \int_{x_0}^{x_1} \sqrt{Q(x,E)} \text{d} x \right). 
\end{align}
In this form, it is valid for energies below the peak of the barrier $E < V_\text{max}$, and becomes less accurate close to it. Although more accurate WKB-based results exist, the advantage of Eq.~\eqref{Gamow} is that it can be inverted similarly to the Bohr-Sommerfeld rule~\eqref{cBS}. In Refs.~\cite{1980AmJPh..48..432L,2006AmJPh..74..638G}, it was shown that it is again the separation of turning points,
\begin{align}\label{widthbarrier}
L_2(E) &= x_1(E)-x_0(E) 
\\
&=  \frac{1}{\pi} \int_{E}^{E_\text{max}} \frac{\left(\partial T(E^\prime)/ \partial E^\prime \right)}{T(E^\prime) \sqrt{E^\prime -E}} \text{d} E^\prime,
\end{align}
which is the universal feature defining a family of potentials with the same transmission $T(E)$. Here $E_\text{max}$ is the value of the potential at its maximum. One can approximate it by solving $T(E_\text{max})=1/2$.

\subsubsection{Parabolic approximation of the potential maximum}

Because the Gamow formula and its inversion formula become less accurate around the potential peak, we approximate this region of the reconstructed, energy-independent, potential barriers with a parabolic approximation
\begin{align}\label{parabollicfitting}
V_{\text{parabolic}}(x)=V_\mathrm{max}+\alpha(x-x_\mathrm{max})^{2},
\end{align}
with $x_\text{max}=0$. The two other free parameters ($V_\mathrm{max}$ and $\alpha$) are then obtained by fitting the analytic form of the transmission to the numerical transmission obtained from the energy-dependent potentials; see Ref.~\cite{Albuquerque:2023lzw}, and the Appendix of Ref.~\cite{Volkel:2019ahb} for more details. In general, we note that the value of $V_\mathrm{max}$ obtained by this fitting procedure agrees well with $E_\text{max}$ obtained by solving $T(E_{\rm max})=1/2$. Once we have estimated the two parabolic parameters, we match our inverse potential barrier constructed with the inverse Gamow formula with the fitted parabola around the peak via a smooth version of the Heaviside function, given by $\Theta_{\pm}(E;E_{0},\kappa)=1/2\pm(1/2)\tanh[\kappa (E-E_{0})]$. Here, $\kappa$ controls the ``smoothness'' of the transition between the two connected curves, while $E_{0}$ determines around where the transition takes place. We choose $E_{0}$ as the energy for which the two curves intersect.

\subsection{Remark on turning points}\label{methods_turning}

Turning points of a potential, as shown in Fig.~\ref{parabollicsimple}, are a concept that is well-motivated beyond WKB theory. Because they play an important role in the expected validity of the WKB method [via $Q(x,E)\equiv E-V(x,E)$], it is natural to generalize the definition of turning points $x_i$ for energy-dependent potentials to $V(x_i,E) = E$, where the argument in $V(x,E)$ is the same $E$ as used on the right-hand side. 

The minimum $V_\text{min}$ or maximum $V_\text{max}$ of an energy-independent, two turning point potential can be defined as the pair of values $(x_\text{critic}, E_\text{critic})$ for which $Q(x,E)=0 \Leftrightarrow V(x)=E$ and $\text{d}Q(x,E)/\text{d}x=0 \Leftrightarrow \text{d}V(x)/\text{dx}=0 $. Similarly, for the more general case of energy-dependent potentials $V(x,E)$, we define the vertex point as the pair of values  $(x_\text{vertex}, E_\text{vertex})$ for which $Q(x_\text{vertex}, E_\text{vertex})=0$, and $\text{d} Q (x_\text{vertex}, E_\text{vertex})/\text{d}x=0$.
Note that another way of defining $V_\text{min}$ or $V_\text{max}$ of a two turning point, energy-independent potential is by requesting that the left and right turning points converge $x_0(E_\text{vertex}) = x_1(E_\text{vertex})$. 
For energy-dependent potentials, this holds as well.

Finally, the Bohr-Sommerfeld treatment of complex-valued potentials, e.g., as they appear for perturbations of the Kerr black hole, is also possible and has been investigated in Ref.~\cite{Kokkotas:1991vz}. In this case, the turning points are, in general, also complex-valued, but both cases share a nontrivial behavior of $Q(x,E)$ with respect to $E$.

\subsection{Numerical methods}\label{methods_NU}

To verify the accuracy of the WKB-based methods, we use a straight-forward shooting method to compute the bound states and numerical integration through the potential barrier to obtain the transmission. Later, we also present the approximate results obtained with the Bohr-Sommerfeld rule and Gamow formula. Such a comparison allows us to determine where the methods are reliable and where they lose accuracy.

\subsubsection{Shooting method for bound states}\label{methods_NU_sh}

The shooting method is based on numerically integrating the wave-equation for given boundary conditions from both sides. For Eq.~\eqref{wave_equation}, we choose the asymptotic behavior for the solution to be given by 
\begin{align}\label{asymptotic_behavior1}
\psi_1 &\rightarrow e^{ \left(\sqrt{-Q(x,E)} x \right)}, & x \rightarrow -\infty,
\\
\psi_2 &\rightarrow e^{\left( -\sqrt{-Q(x,E)} x\right)}, & x \rightarrow \infty.
\end{align}
for the cases where $V(x,E)$ asymptotically goes to constant values, and for the cases where $V(x,E)$ diverges at the two limits.  
The eigenvalues are then obtained by determining the roots of the Wronskian of the numerical solutions at some intermediate point. The Wronskian of the two solutions is defined as
\begin{equation}
W(x,E;\psi_{1},\psi_{2}) \equiv \psi_{1}(x)\psi_{2}^{\prime}(x)-\psi_{2}(x)\psi_{1}^{\prime}(x).
\end{equation}

\subsubsection{Numerical integration for transmission}

Similar to the shooting method, the transmission is obtained by numerical integration, but for different boundary conditions. In this case, our ansatz for two independent solutions satisfy the following boundary conditions
\begin{align}\label{transmitionsolution1}
\psi_{1} \rightarrow
\begin{cases}
e^{\left(- \mathrm{i} \sqrt{Q(x,E)} x \right)},&   x\rightarrow - \infty, \\  A^{-}_{\infty}e^{\left(- \mathrm{i} \sqrt{Q(x,E)} x \right)} \\
\qquad + A^{+}_{\infty}e^{\left(+ \mathrm{i} \sqrt{Q(x,E)} x \right)}, &  x\rightarrow + \infty, 
\end{cases}
\end{align}
and
\begin{align}\label{transmitionsolution2}
\psi_{2}\rightarrow 
\begin{cases}
B^{-}_{-\infty}e^{\left(- \mathrm{i} \sqrt{Q(x,E)} x \right)} \\
\qquad + B^{+}_{-\infty}e^{\left(+ \mathrm{i} \sqrt{Q(x,E)} x \right)}, &   x\rightarrow - \infty, \\  e^{\left( \mathrm{i} \sqrt{Q(x,E)} x \right)} . &  x\rightarrow + \infty. 
\end{cases}
\end{align}
From the computational point of view, the solutions described by $\psi_{1}$ and $\psi_{2}$ describe monochromatic plane waves numerically evolved from one end of the domain to the other. Physically, $\psi_{1}$ represents incoming plane waves with an incident amplitude $A^{+}_{\infty}$ that are scattered at the energy-dependent potential $V(x,E)$. Those waves are partially reflected with reflection amplitude $A^{-}_{\infty}/A^{+}_{\infty}$, and partially transmitted with transmission amplitude $1/A^{-}_{\infty}$. A similar interpretation can be made for $\psi_{2}$. Accordingly, the transmission coefficient can be defined as
\begin{align}\label{transmitioncoefficient}
\abs{T}^2=\frac{1}{\abs{A^{-}_{\infty}}^2}=\frac{1}{\abs{B^{-}_{-\infty}}^2}.
\end{align}

\section{Application and results}\label{app_results}

In the following, we apply the WKB-based methods to two types of energy-dependent potentials. In Sec.~\ref{app_parabola}, we study a modified quadratic potential, and in Sec.~\ref{app_PT}, we consider a modified P\"oschl-Teller potential. We first consider parameters of the potentials that yield wells admitting bound states and then barriers for which we compute the transmissions.  

In all applications, our approach can be explained in three main steps. First, for the given energy-dependent potential, we provide the spectrum of bound states/transmission either analytically or with the numerical method. Second, we use the spectrum/transmission as input for the WKB-based inverse method to construct the width of a family of WKB-equivalent, energy-independent potentials. Third, we use the numerical method to compute the bound states/transmission of one of the reconstructed potentials  $V_{\text{inv}}(x)$ to quantify how accurately they match the original ones. For comparison, we also use the numerical method to compute the associated properties of the energy-independent, width-equivalent potential $V_\text{width}(x)$.

Therefore, we associate with a certain energy-dependent potential $V(x,E)$, an energy-independent, inverse potential $V_{\text{inv}}(x)$, which must match the property used for its reconstruction, at least within the validity of the WKB-approximation. This allows one to describe properties of energy-dependent potentials more simply by constructing energy-independent ones.

\subsection{Energy-dependent quadratic potential}\label{app_parabola}

As first example, we study an energy-dependent quadratic potential, given by 
\begin{align}\label{parabollicdependent}
V_\text{HO}(x,E) = (a+bE)x^2+cE+V_{0},
\end{align}
where $a,b,c,V_0$ are real-valued constants. In the following, we consider two scenarios that either describe potential wells in Sec.~\ref{app_ho_wells}, or potential barriers in Sec.~\ref{app_ho_barriers}. For each of the two scenarios, we study two distinct cases for the variation of the concavity with the variation of the energy. The cases are defined by $a/b<0$, which tends to increasingly open the parabola with increasing $E$, and by $a/b>0$, which tends to close it. To make the main text more readable, but still cover comprehensive material, some of the results are reported in the Appendix~\ref{appendix}.

\subsubsection{Potential wells}\label{app_ho_wells}

The spectrum of bound states $E_n$ can be computed analytically. For this we generalized results first reported for the case $c=V_0=0$ in Ref.~\cite{Form_nek_2004}. The energy of the normal modes are obtained by solving
\begin{align}\label{HO_En}
E_n (1-c)+V_{0} = (2n+1)\sqrt{a+b E_n}.
\end{align}
To test our numerical implementation of the shooting method presented in Sec.~\ref{methods_NU_sh}, we have  verified that it agrees with the analytic results. 

\begin{figure}
\includegraphics[width=1.0\linewidth]{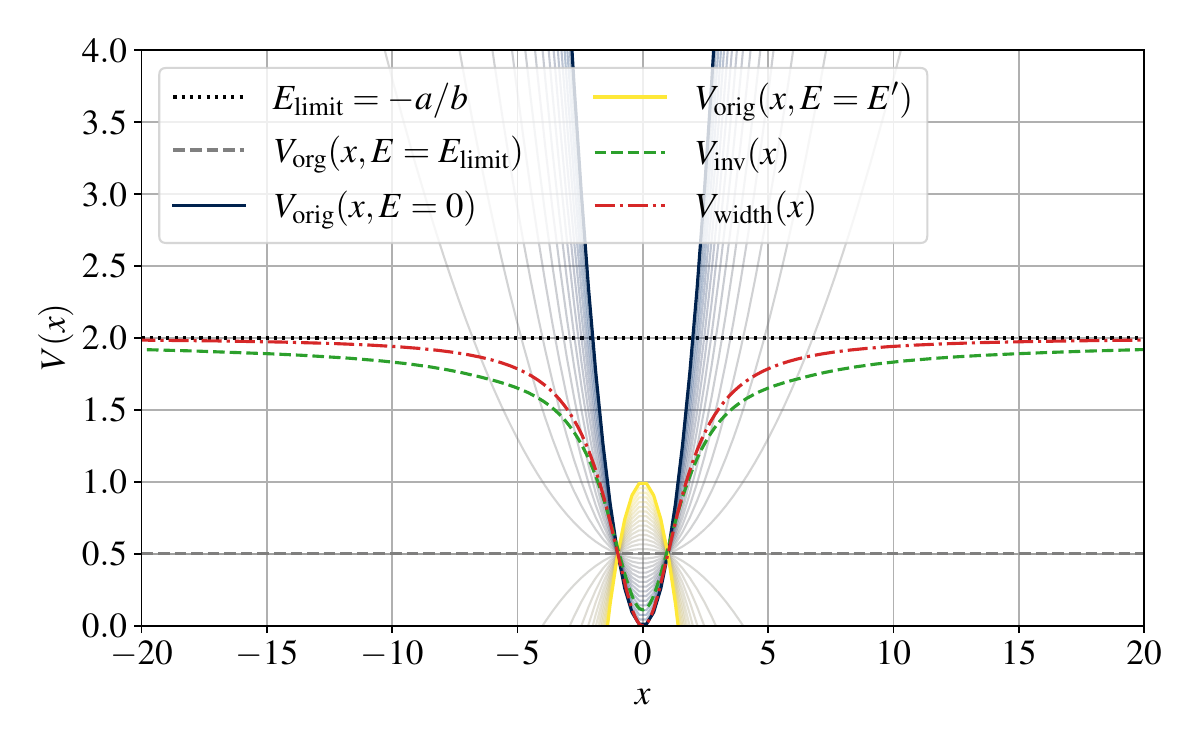}
\includegraphics[width=1.0\linewidth]{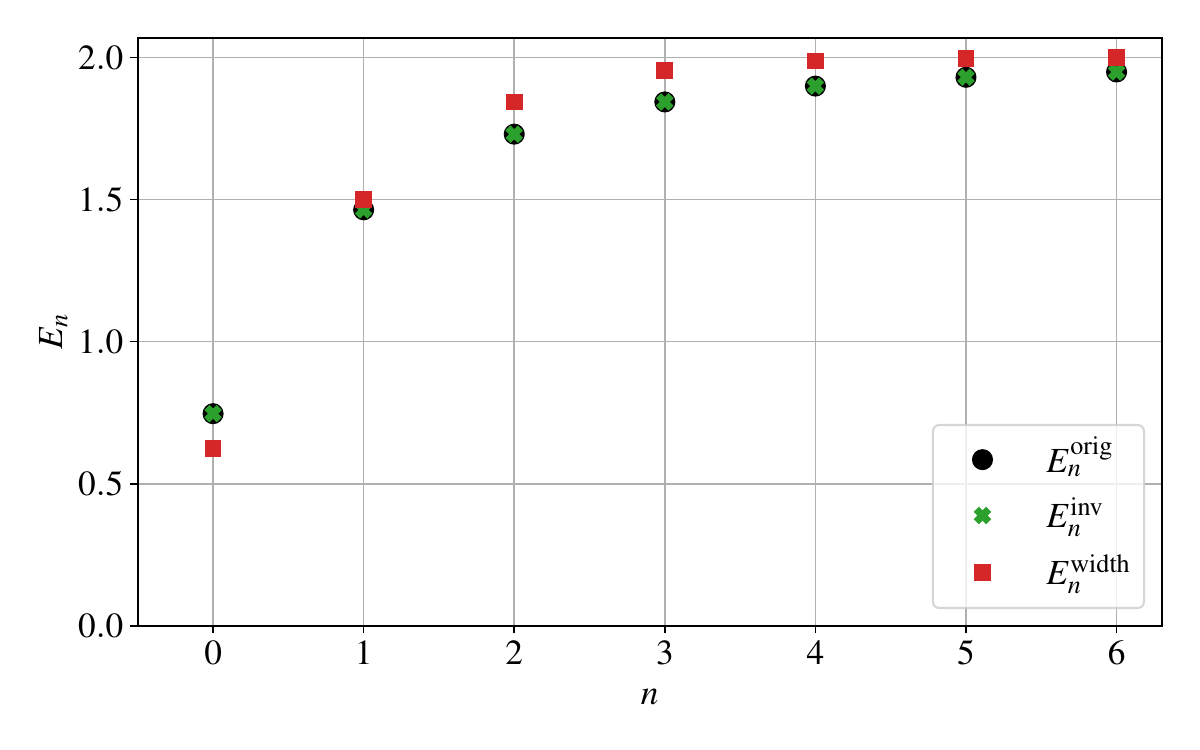}
\caption{
Application of the inverse method to the energy-dependent quadratic potential Eq.~\eqref{parabollicdependent} for case $a/b<0$, with parameters $a=0.5, b=-0.25, c=0.25, V_{0}=0$. Top panel: The series of curves of varying colors indicate the chosen value of $E$ when computing the energy-dependent potential $V_\text{orig}(x, E)=V_\text{HO}(x,E)$ starting from $E=0$ to $E=E^\prime= 2 E_{\text{limit}}$. The inverse potential (green dashed) is labeled as $V_\text{inv}(x)$ and the width-equivalent potential (red dotted dashed) is labeled as $V_\text{width}(x)$. Bottom panel: Here we show the spectrum of bound states for the original potential $E^\text{orig}_n$ (black circles), the inverse potential $E^\text{inv}_n$ (green cross), and the width-equivalent potential $E^\text{width}_n$ (red squares). 
\label{parabollicresults}
}
\end{figure}

Varying $E$ yields a family of curves for the potential, whose qualitative properties depends nontrivially on the chosen parameters. To illustrate that, we present the potential and the results of the inverse method for $a/b<0$ in Fig.~\ref{parabollicresults}, whose caption contains the numerical values of all parameters. The top panel demonstrates that the potential curves as function of $E$ get increasingly more open until $E_\text{limit}=-a/b$. At this energy, the potential turns into a horizontal line at $V_{0}-ca/b$ and the separation of turning points diverges. For even higher energies, the curvature of the parabola becomes negative. Qualitatively, we therefore expect that by approaching the energy limit of $E\rightarrow E_\text{limit} = -a/b$, the system starts behaving as a quasifree particle in a constant potential, with the energy spacing of the modes becoming increasingly smaller. 

Using the bound states as input for the inverse Bohr-Sommerfeld rule~\eqref{Excursion} yields a family of inverse potentials $V_\text{inv}(x)$ sharing the same separation of turning-points. To fix one of them, we assume that the potential is symmetric around the origin ($x_0(E)=-x_1(E)$). From Fig.~\ref{parabollicresults} it is evident that $V_\text{inv}(x)$ is not the same as the width equivalent potential $V_\text{width}(x)$. On the other hand, they are asymptotically converging to the same maximum energy for large values of $x$ and are similar close to the minimum. 

In the bottom panel of Fig.~\ref{parabollicresults}, we compare the bound states of the original potential Eq.~\eqref{HO_En} with the ones of the inverse potential and the width-equivalent potential, all computed using the numerical method. The inverse potential $V_\text{inv}(x)$ is quasi-isospectral with the original energy-dependent potential. The asymptotic behavior of the overtones marks the transition from an essentially discrete spectrum to a quasicontinuum one, showing the increasing opening tendency of the energy-dependent potential. The bound states of the width-equivalent potential $V_\text{width}(x)$ differ quantitatively but otherwise have a similar behavior as a function of $n$.

The case $a/b>0$ also provides interesting, but qualitatively different applications. Instead of opening with higher energy values, the potential curves become progressively more closed. We refer the reader to Appendix \ref{app1} for an illustration of this scenario.

\subsubsection{Potential barriers}\label{app_ho_barriers}

To represent a more standard barrier, we redefine the original potential as follows
\begin{align}\label{potentialtransmissionquadratic}
\bar{V}_\text{HO}(x,E)=\biggl\{\begin{array}{ll} V_{\rm HO}(x,E),&   x_{i 1}< x< x_{i 2}, \\  0, &   \text{otherwise}. \end{array}
\end{align}
The points $x_{i 1}$ and $x_{i 2}$ are defined by where the potential is zero, by $V_{\rm HO}(x_{i1, i2},E)=0$, and are thus given by
\begin{align}\label{intersectionpoints}
    x_{i1,i2}=\pm \sqrt{-\frac{V_{0}+cE}{a+bE}}.
\end{align}

As previously, the energy dependence introduces a nontrivial behavior for the family of potential curves. The linear term $cE$ is responsible for the vertical shift. For $c\in \left(0,1\right)$, by increasing the energy, it will eventually reach a vertex point $E_{\rm vertex}=V_{0}/(1-c)$ in which the two turning points converge to a single point at the maximum of its associated parabola $V(x_{\rm vertex},E_{\rm vertex})$, just like a global maximum in an energy-independent potential. For $c>1$, however, this scenario does not occur.

The maximum of an energy-independent potential barrier with two turning points plays an important role in the scattering of waves~\cite{Schutz:1985km}. There the transmission changes from exponentially small values to asymptotically one. Similarly, for energy-dependent potentials, the vertex $E_{\rm vertex}$ for the turning points is exactly where the wave scattering of those potentials transits from almost null absorption to full transmission. After this reference value, the associated potential $V(x,E)$ no longer possesses turning points, and the energy is above the barrier. Thus $E_{\rm vertex}$ characterizes a local maximum of an effective, energy-independent potential barrier with similar properties. The existence of  $E_{\rm vertex}$, which plays the role of $E_\text{max}$ in the Gamow formula, is crucial for the inverse method when applied to energy-dependent potentials. For this reason, our discussions are limited to $c \in \left( 0,1 \right)$. 

\begin{figure}
\includegraphics[width=1.0\linewidth]{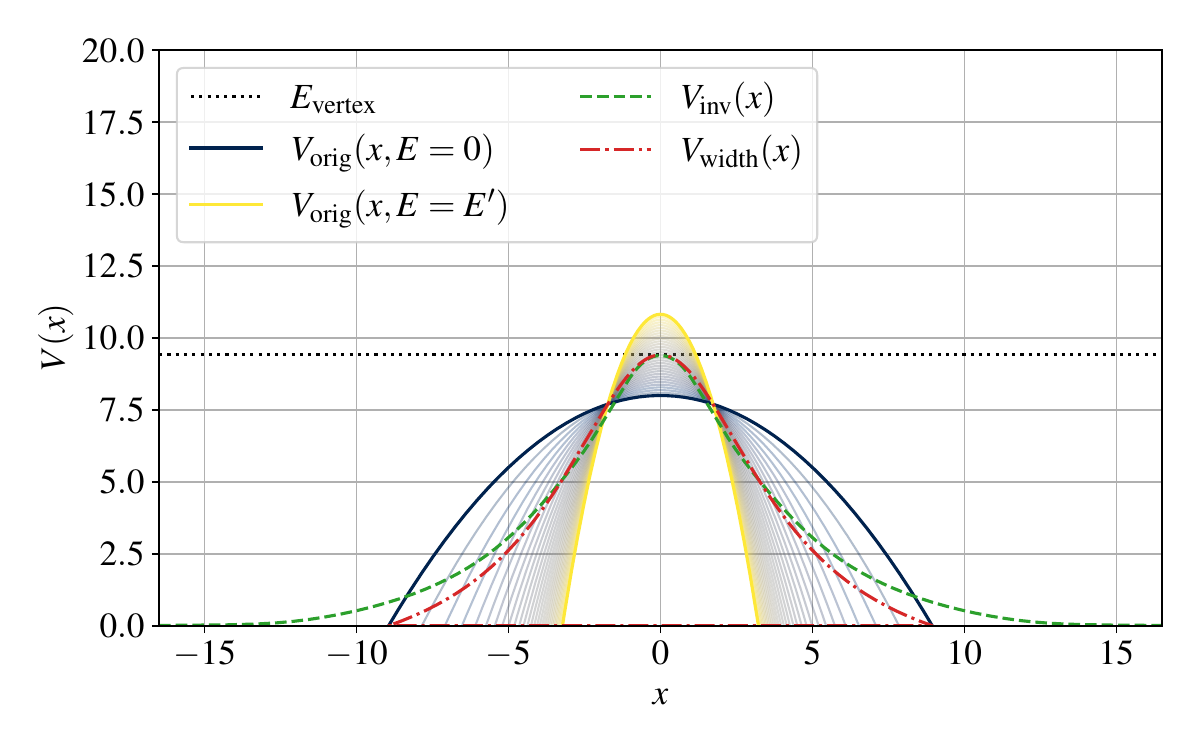}
\includegraphics[width=1.0\linewidth]{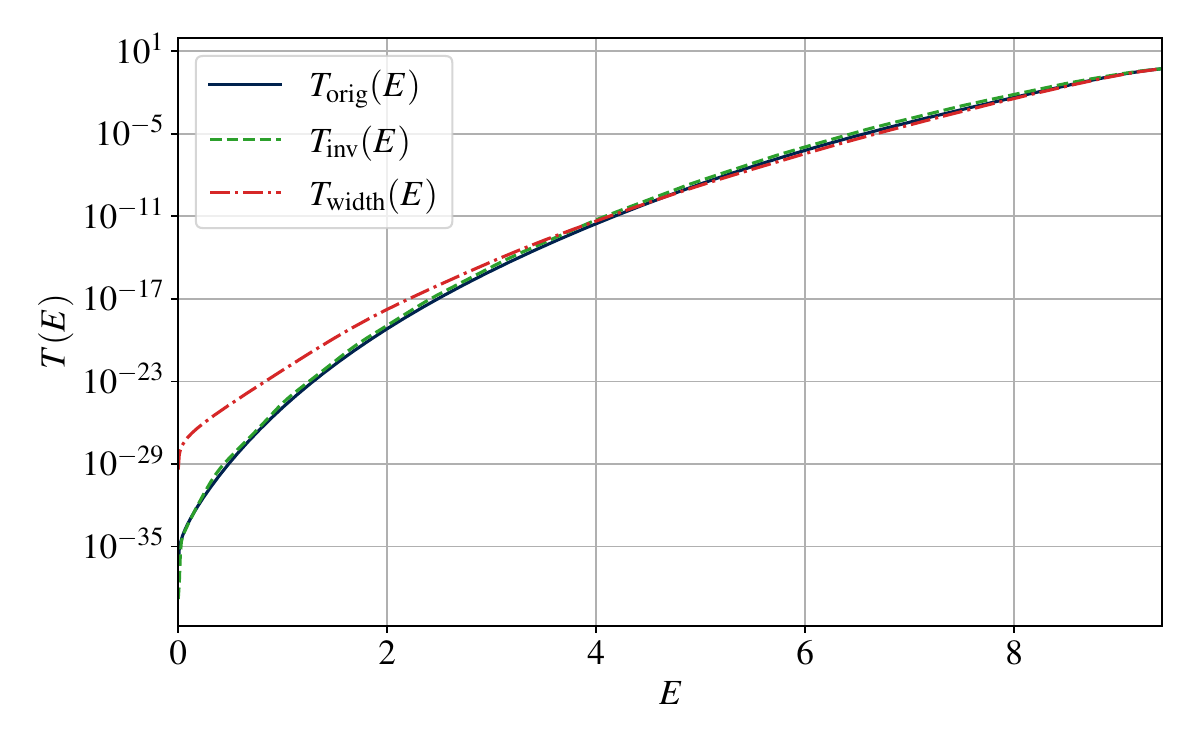}
\caption{ Application of the inverse method to the energy-dependent quadratic potential barrier  Eq.~\eqref{parabollicdependent} for case $a/b>0$, with parameters $a = -0.1$, $b =-0.05$, $c=0.15$, and $V_{0}=8$. Top panel: The series of curves of varying colors indicate the chosen value of $E$ when computing the energy-dependent potential $V_\text{orig}(x, E)=\bar{V}_\text{HO}(x,E)$ starting from $E=0$ to $E=E^\prime= 2 E_{\rm vertex}$. The inverse potential (green dashed) is labeled as $V_\text{inv}(x)$ and the width equivalent potential (red dotted dashed) is labeled as $V_\text{width}(x)$. Bottom panel: Here we show the transmission for the original potential $T_{\rm orig}(E)$ (black solid line), the inverse potential  $T_{\rm inv}(E)$ (green dashed line) and the width-equivalent potential  $T_{\rm width}(E)$ (red dot-dashed lines). \label{parabollicresults_T2}}
\end{figure}

In the following, we apply our method to a potential with $a/b>0$, and report our results in Fig.~\ref{parabollicresults_T2}. It shows that the potential barrier curves tend to increasingly close their concavity with increasing $E$, which is shown in the top panel of Fig.~\ref{parabollicresults_T2}. Here we also provide the results for the inverse potential $V_\text{inv}(x)$ and the width-equivalent potential $V_\text{width}(x)$. They are both similar close to their vertex at $E_{\rm vertex}=V_{0}/(1-c)$, but overall deviate. The corresponding transmissions provided in the bottom panel demonstrate that $V_\text{inv}(x)$ better reproduces the original transmission. The transmission of $V_\text{width}(x)$ differs substantially for small energies, but becomes very similar around the maximum. In the Appendix \ref{app1}, we provide complementary results for a potential barrier with $a/b<0$.

\subsection{Energy-dependent P\"oschl-Teller potential}\label{app_PT}

As the second main example of energy-dependent potentials, we introduce a modified P\"oschl-Teller potential given as follows
\begin{align}\label{ptdefinition}
V_\text{PT}(x,E) = (a+bE)\text{sech}^2[k(x-x_{0})]+cE + V_{0}.
\end{align}
The main reason for considering the P\"oschl-Teller potentials is that it is widely used in different areas of physics, including nuclear physics \cite{ptnuclear1,ptnuclear2}, and the perturbations of black holes and exotic compact objects~\cite{Mashhoon:1982im,Ferrari:1984ozr,Price:2017cjr}. 

An important difference between the energy-dependent, quadratic potential and the here presented  P\"oschl-Teller potential is that the latter one converges necessarily to finite asymptotic values for $x\rightarrow \pm \infty$ given by $V_{0}+cE$.

\subsubsection{Potential wells}

\begin{figure}
\includegraphics[width=1.0\linewidth]{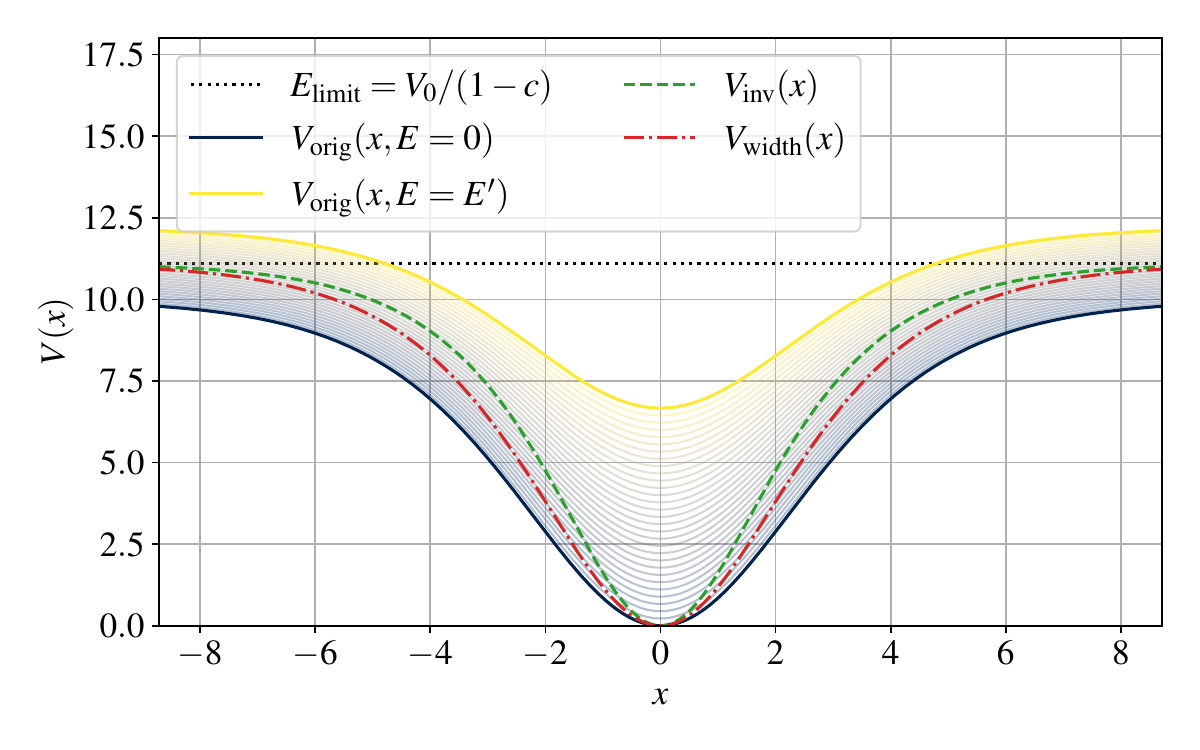}
\includegraphics[width=1.0\linewidth]{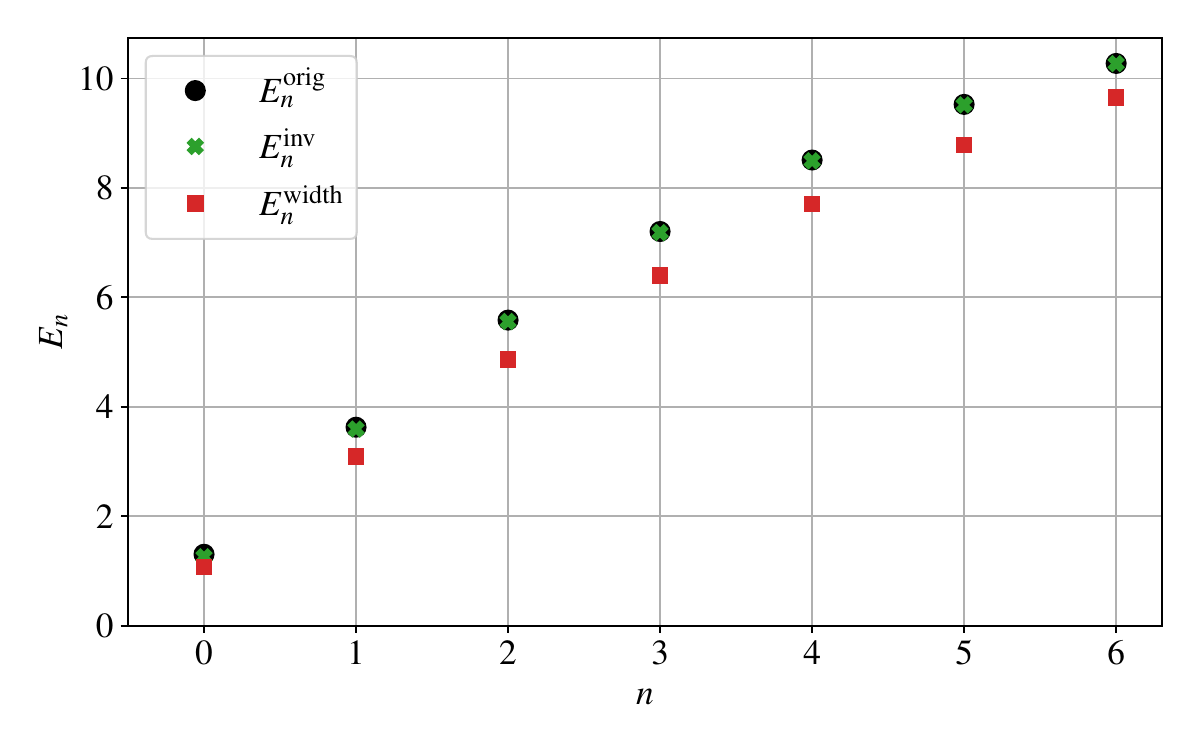}
\caption{ Application of the inverse method to the energy-dependent P\"oschl-Teller potential Eq.~\eqref{ptdefinition} for case $a/b<0$, with parameters $a = -10$, $b = 0.2$, $c=0.1$, $k=0.3$, and $V_{0}=10$. Top panel: The series of curves of varying colors indicate the chosen value of $E$ when computing the energy-dependent potential $V_\text{orig}(x, E)=V_\text{PT}(x,E)$ starting from $E=0$ to $E=E^\prime= 2 E_\text{limit}$. The inverse potential (green dashed) is labeled as $V_\text{inv}(x)$ and the width equivalent potential (red dotted dashed) is labeled as $V_\text{width}(x)$. Bottom panel: Here we show the spectrum of bound states for the original potential $E^\text{orig}_n$ (black circles), the inverse potential $E^\text{inv}_n$ (green cross), and width-equivalent potential $E^\text{width}_n$ (red squares).  \label{ptresults_En_PT1}}
\end{figure}

In this subsection, we present the findings of the P\"oschl-Teller energy-dependent potential wells. Again, we differentiate between $a/b>0$ and $a/b<0$. Both cases provide distinct profiles for the energy-dependent variation of the potential. We show the potentials for a range of energy values for $a/b<0$ in the top panel of Fig.~\ref{ptresults_En_PT1}. 
Here, we also present the inverse potential and the width-equivalent potential. In the bottom panel, we report the associated bound states. The case $a/b>0$ is presented in Appendix \ref{appendix}.

We notice that $V_\text{inv}(x)$ and $V_\text{width}(x)$ agree in both cases very well at the minimum, as well as for their asymptotic value $V_{0}/(1-c)$ for large values of $|x|$. However, for intermediate energies, the two potentials differ. The bound states of the original potential agree in both cases very well with the ones of the inverse potential, while they differ for the ones of the width-equivalent one. As the energy approaches the asymptotic value of the potential, the energy spacing between the modes decreases, which indicates the transition to a quasicontinuum spectrum of a quasifree particle, as in Sec.~\ref{app_ho_wells}. Note that the energy dependence lifts the potential minimum, and thus decreases the space for bound states. The opposite is observed in case $a/b>0$, which is shown in Appendix \ref{app2}. 

\subsubsection{Potential barriers}

\begin{figure}
\includegraphics[width=1.0\linewidth]{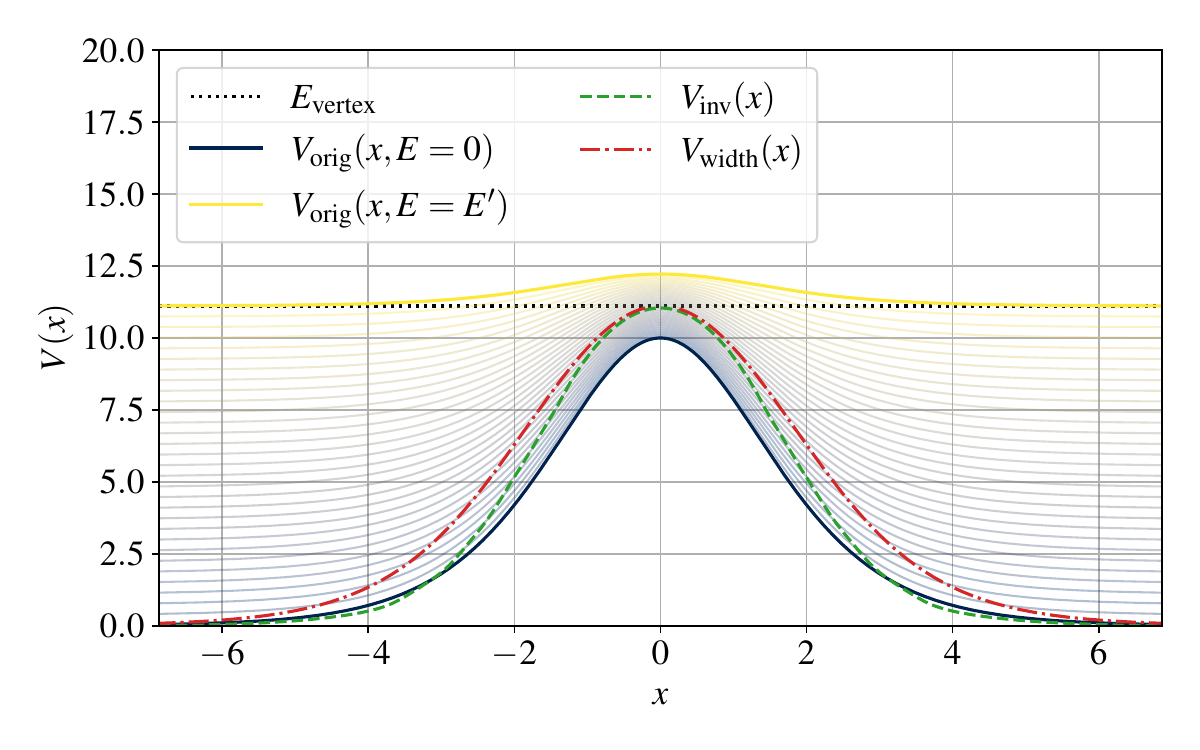}
\includegraphics[width=1.0\linewidth]{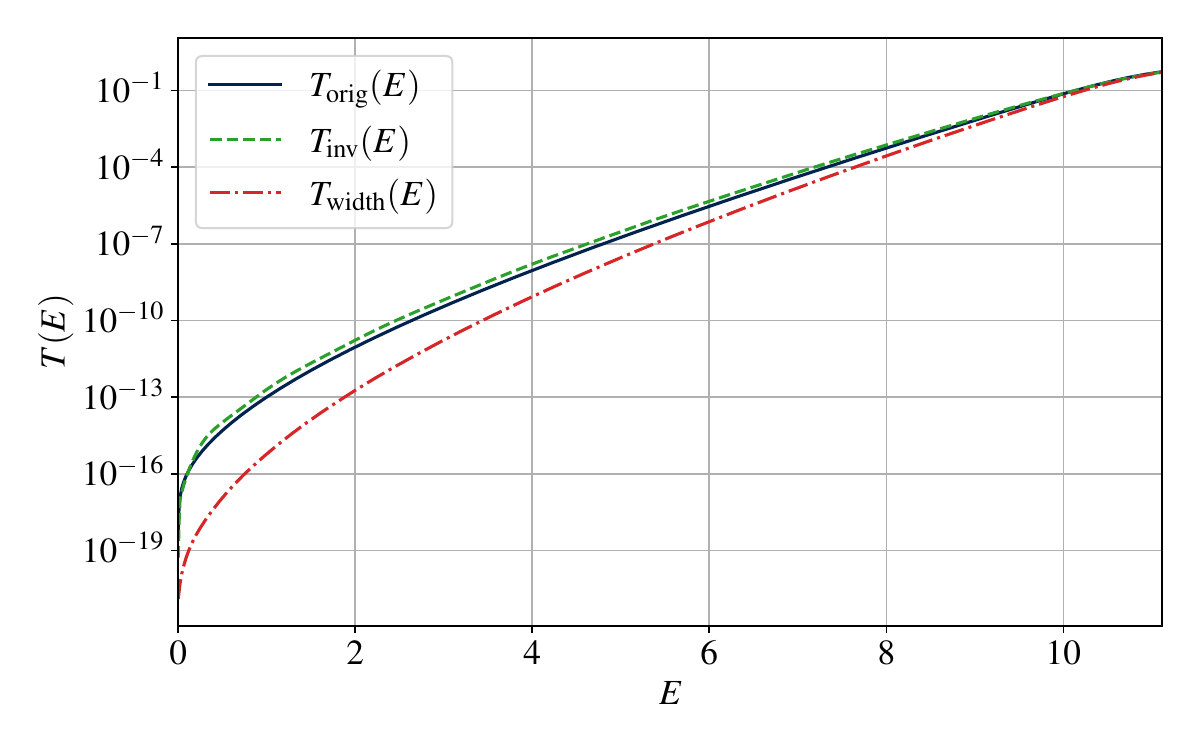}
\caption{
Application of the inverse method to the energy-dependent, P\"oschl-Teller potential Eq.~\eqref{parabollicdependent} for case $a/b<0$ with parameters $a=10$, $b=-0.4$, $c=0.5$, $k=0.5$, and $V_{0}=0$. Top panel: The series of curves of varying colors indicate the chosen value of $E$ when computing the energy-dependent potential $V_\text{orig}(x, E)=V_\text{PT}(x,E)$ starting from $E=0$ to $E=E^\prime=  2 E_\text{vertex}$. The inverse potential (green dashed) is labeled as $V_\text{inv}(x)$ and the width equivalent potential (red dotted dashed) is labeled as $V_\text{width}(x)$. Bottom panel: Here we show the transmission for the original potential $T_{\rm orig}(E)$ (black solid line), the inverse potential  $T_{\rm inv}(E)$ (green dashed line) and the width-equivalent potential  $T_{\rm width}(E)$ (red dot-dashed lines). \label{ptresults_TE_PT1} }
\end{figure}

Finally, we investigate P\"oschl-Teller potential barriers. In the following, we consider $a/b<0$, while $a/b>0$ is reported in Appendix \ref{appendix}. The potentials can be found in the top panel of Fig.~\ref{ptresults_TE_PT1}, while the associated transmissions are reported in the bottom panel. In both cases, we find that $V_\text{inv}(x)$ and $V_\text{width}(x)$ agree well at their maximum value $E_\text{vertex}=(a+V_{0})/(1-(c+b))$ , but not at intermediate energies.  In the lower panel, we show the transmissions associated with the potentials. As expected, the transmission reconstructed from the inverse potential matches very well with the original transmission, while there are significant differences when compared to the width-equivalent potential, at least for energies below the maximum.

\section{Discussion}\label{discussion}

\subsection{Width-equivalent is not WKB-equivalent}

As stated previously, and demonstrated in our results, one key finding of our work is that the spectra and transmissions of width-equivalent energy-dependent potentials are, in general, not WKB-equivalent. By explicitly constructing the width-equivalent potentials and computing their spectral properties with an accurate numerical method, it is evident that they no longer correspond to those of the energy-dependent potential. 

For the bound states $E_n$, we noticed that the deviations were typically small around $n=0$ and then increased. One can explain this behavior by observing that the width equivalent and inverse potentials agree well around their minimum. For the transmission $T(E)$, one finds that the agreement between both potentials is good around the maximum; also a consequence of the local approximation of the peak, and then deviates for smaller energies. In both cases, the spectral properties can be well-understood from the local character of the Bohr-Sommerfeld rule and Gamow formula.

\subsection{Accuracy of inverse methods}

Because the WKB method is in general not exact, using the Bohr-Sommerfeld rule and Gamow formula for the direct problem, or their inversions for the inverse problem, can in general only provide approximate results. This is well known for energy-independent potentials and it also holds for the more general, energy-dependent ones. To demonstrate this, we vary the potential properties of some of the previous cases to further investigate the accuracy of the WKB method. We compare the WKB predictions of the bound states $E_n$ and transmission $T(E)$ of the original potential, with the ones of the WKB-constructed inverse potential predicted using the accurate numerical method. For the overall performance of the WKB method for the inverse potential, as one would expect from the direct problem, the spectral properties of the inverse potentials should match the ones of the original potential more accurately for higher bound states and energies below the peak of the barrier.

\begin{figure}
\includegraphics[width=1.0\linewidth]{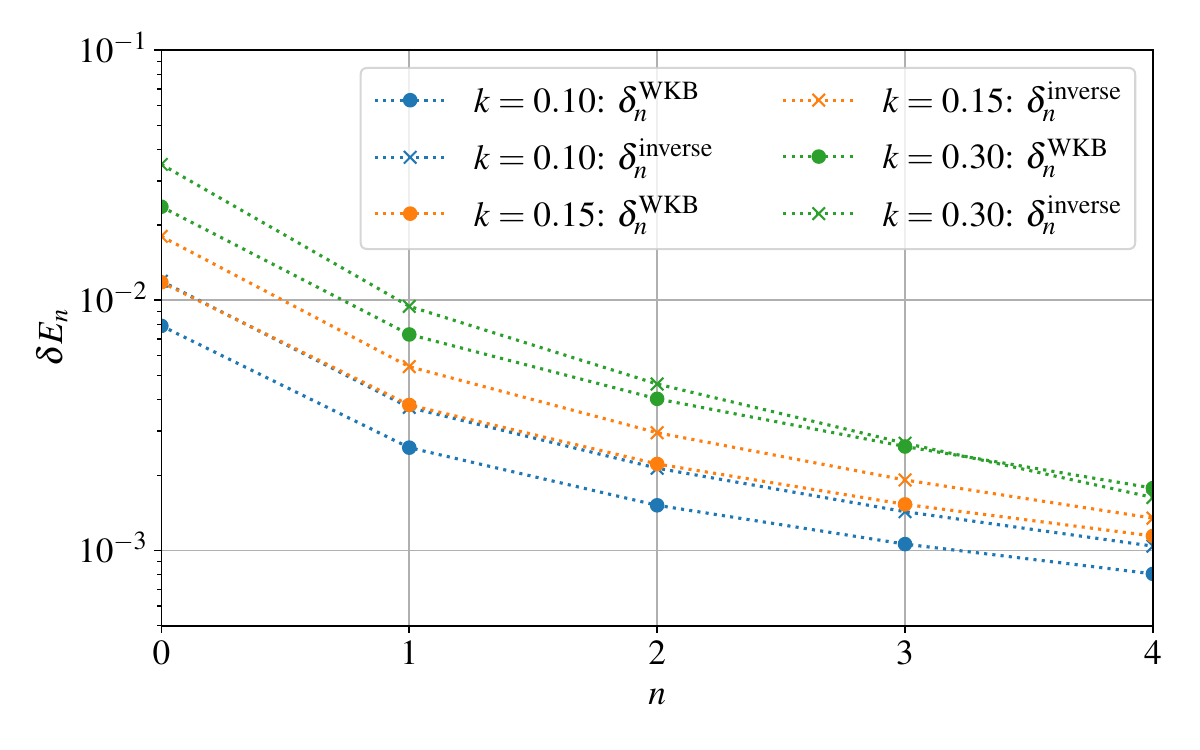}
\caption{
Here we show the relative error defined by $\delta^{i}_{n}$, where the index $i$ stands for either the WKB prediction for the original potential (circles) or the numerical method of the inverse potential (crosses). The different colors represent different choices of $k$, which changes the width of the well, and thus indirectly the expected accuracy of the WKB method. All other parameters are those of Fig.~\ref{ptresults_En_PT1}. 
 \label{relativeerror1} 
 }
\end{figure}

In Fig.~\ref{relativeerror1}, we computed the relative errors of the bound states via
\begin{align}
\delta^{i}_{n}  \equiv \left| \frac{E^\text{Num orig}_n-E^\mathrm{i}_n}{E^\text{Num orig}_n} \right|,
\end{align}
where $i$ is either using the direct WKB prediction for the original potential or the numerical prediction for the inverse potential. For increasing bound state number $n$, we report that the relative errors decrease, and that the performance of the inverse potential is clearly correlated with the accuracy of the WKB method for the direct problem. We also find that in most cases, the relative errors of the direct WKB computation of the original potential are smaller than those of the numerical method of the inverse potential. Since the construction of the inverse potential requires one to interpolate the spectrum of bound states, an overall difference in accuracy should be expected from this additional source of imprecision. 

\begin{figure}
\includegraphics[width=1.0\linewidth]{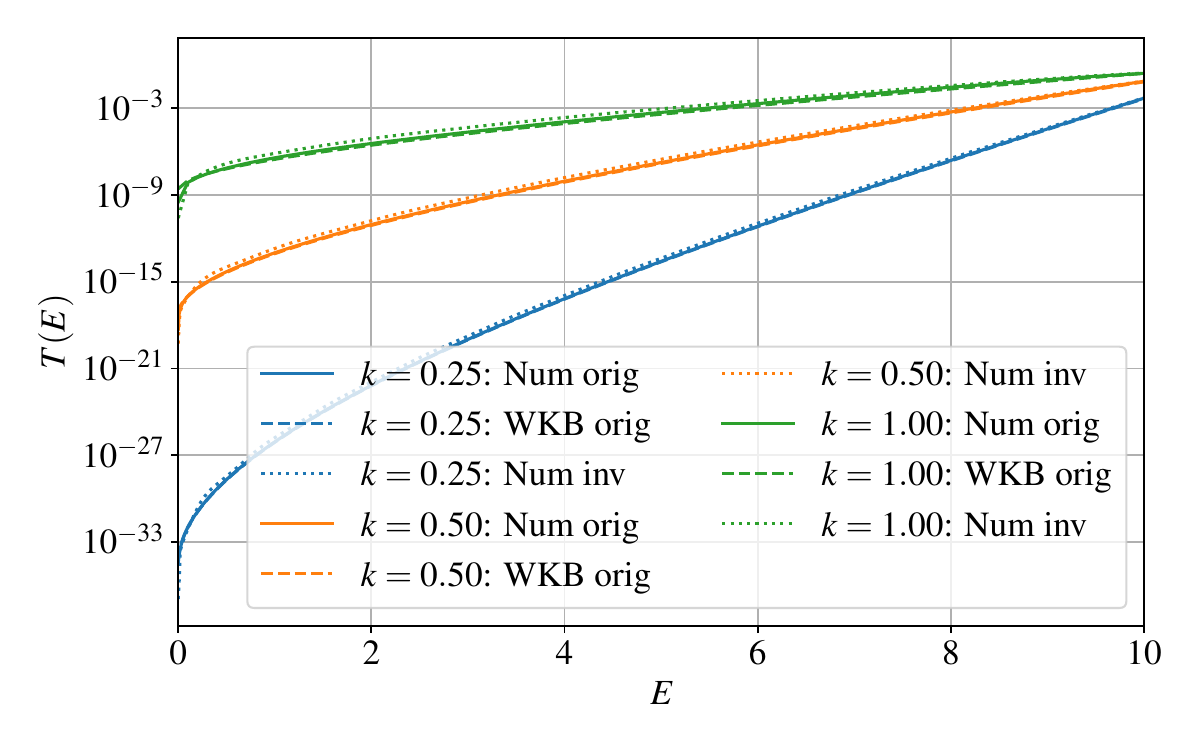}
\caption{
Here we show the transmission $T(E)$ using the numerical method for the original potential (solid lines), the WKB prediction for the original potential (dashed lines), and the numerical method of the inverse potential (dotted lines). The different colors represent different choices of $k$, which changes the width of the barrier, and thus indirectly the expected accuracy of the WKB method. All other parameters are those of Fig.~\ref{ptresults_TE_PT1}.
 \label{relativeerror2} }
\end{figure}

Next we investigate the performance when computing the transmission $T(E)$, which is shown in Fig.~\ref{relativeerror2}. As previously, we vary $k$ of one of the previous cases and leave all other parameters to be the same. Our results confirm what should be expected. For wider potential barriers (smaller values of $k$), the different predictions are more similar throughout all energies. Since the inverse method does not rely on an interpolation of the transmission, both WKB related predictions should be of similar accuracy with respect to the numerical result of the original potential. 

We conclude this part of the discussion by noting that the good agreement between results of the numerical original potential and the numerical inverse potential are also an independent check of the numerical method itself. 

\subsection{Remarks on nonuniqueness}

Even without energy-dependence, the inverse problem is in general not uniquely solvable. The WKB-equivalent potentials constructed from the inverse, semiclassical methods provide an intuitively clear, but approximate answer. They map a one-dimensional function of energy, $n(E)$ or $T(E)$, into the width $L(E)$ of all possible potential wells or barriers. Note that it might be possible to provide one turning point function from underlying properties of the application, which would then uniquely determine the underlying potential, e.g., Refs.~\cite{Volkel:2017kfj,Volkel:2018hwb}.

Introducing energy dependence to $V(x,E)$ does not translate to a family of bound states or transmission functions, but instead, only yields another one-dimensional function of energy. This is fundamentally different from introducing a free parameter, for which a family of bound states or transmissions could be mapped to a family of widths. A closer look into the Bohr-Sommerfeld and Gamow integrals reveals that potentials defined by the separation of turning points via $E=V(x,E)$, are in general not equivalent to the reconstructed potentials, which are instead isospectral to the energy-dependent potential. Future extensions of this work could study to what extend it may be possible to infer the energy-dependent contributions to $V(x,E)$, e.g., if it can be treated perturbatively. 

Finally, the inverse problem could in general be ill-posed, and thus any inverse approach must fail. Independent of the nonuniqueness, a typical situation in which the construction of an inverse potential can fail when using the WKB-based methods, is when the width of the well/barrier is not strictly monotonically increasing/decreasing as function of $E$. In such a case, there are ``overhanging cliffs'' in the potential (see Wheeler~\cite{lieb2015studies}), because there is no bijective mapping from the width to a well-defined potential. For some explicit examples, we refer the interested reader to Ref.~\cite{Volkel:2018czg}.

\section{Conclusions}\label{conclusions}

In this work, we have studied the direct and inverse problem of energy-dependent potentials using results from WKB theory. From the inversion of the classical Bohr-Sommerfeld rule and the Gamow formula for energy-independent potentials, it is well-known in the literature that the reconstructed potentials are not unique. Instead, there is a family of infinitely many potentials that share a common property, which is the separation of their classical turnings points, also known as width. 

Because results of the inverse method have been limited to energy-independent potentials, but many physical applications require energy dependency, we have extended it to the inversion of such potentials. Here we have focused on introducing energy-dependent terms to the quadratic potential (harmonic oscillator) and P\"oschl-Teller potentials, which serve as examples that offer a rich phenomenology. 

By applying the same inversion techniques to the bound states or transmission coefficients of energy-dependent potentials, we have explicitly demonstrated that it is possible to construct a family of energy-independent potentials. We note that the bound states and transmission coefficients used for the inverse method are either known analytically or have been computed with full numerical methods, and are thus not limited to the accuracy of the WKB method. We have also used full numerical methods to verify the accuracy of the reconstructed potentials by computing their bound states and transmission coefficients, and find it is comparable to the expected accuracy of the WKB method for the direct problem. 

The novelty of our findings is that the widths of the reconstructed potentials are not equivalent to the original, energy dependent ones. Thus, energy-dependent WKB equivalent potentials are not width-equivalent anymore. We plan to utilize these findings to extend recent work on the inverse problem of analog gravity systems~\cite{Albuquerque:2023lzw} to make the inversion technique applicable to a broader class of systems. Other possible extensions could include the study of rotating, exotic compact objects, whose nonrotating spectral properties have been used for the inverse problem in Refs.~\cite{Volkel:2017kfj,Volkel:2018hwb}. 

\acknowledgments
S.\,A. acknowledges funding from Conselho Nacional de Desenvolvimento Cient\'ifico e Tecnol\'ogico (CNPQ)-Brazil and Coordena\c{c}\~ao de Aperfei\c{c}oamento de Pessoal de N\'ivel Superior (CAPES)-Brazil. 
S.\,H.\,V. acknowledges funding from the Deutsche Forschungsgemeinschaft (DFG) - Project No. 386119226. 
{K.~D.~K and S.\,H.\,V. acknowledge support from the T\"ubingen-Nottingham Joint Seedcorn Fund.}

\bibliography{literature}

\begin{thebibliography}{46}%
\makeatletter
\providecommand \@ifxundefined [1]{%
 \@ifx{#1\undefined}
}%
\providecommand \@ifnum [1]{%
 \ifnum #1\expandafter \@firstoftwo
 \else \expandafter \@secondoftwo
 \fi
}%
\providecommand \@ifx [1]{%
 \ifx #1\expandafter \@firstoftwo
 \else \expandafter \@secondoftwo
 \fi
}%
\providecommand \natexlab [1]{#1}%
\providecommand \enquote  [1]{``#1''}%
\providecommand \bibnamefont  [1]{#1}%
\providecommand \bibfnamefont [1]{#1}%
\providecommand \citenamefont [1]{#1}%
\providecommand \href@noop [0]{\@secondoftwo}%
\providecommand \href [0]{\begingroup \@sanitize@url \@href}%
\providecommand \@href[1]{\@@startlink{#1}\@@href}%
\providecommand \@@href[1]{\endgroup#1\@@endlink}%
\providecommand \@sanitize@url [0]{\catcode `\\12\catcode `\$12\catcode
  `\&12\catcode `\#12\catcode `\^12\catcode `\_12\catcode `\%12\relax}%
\providecommand \@@startlink[1]{}%
\providecommand \@@endlink[0]{}%
\providecommand \url  [0]{\begingroup\@sanitize@url \@url }%
\providecommand \@url [1]{\endgroup\@href {#1}{\urlprefix }}%
\providecommand \urlprefix  [0]{URL }%
\providecommand \Eprint [0]{\href }%
\providecommand \doibase [0]{https://doi.org/}%
\providecommand \selectlanguage [0]{\@gobble}%
\providecommand \bibinfo  [0]{\@secondoftwo}%
\providecommand \bibfield  [0]{\@secondoftwo}%
\providecommand \translation [1]{[#1]}%
\providecommand \BibitemOpen [0]{}%
\providecommand \bibitemStop [0]{}%
\providecommand \bibitemNoStop [0]{.\EOS\space}%
\providecommand \EOS [0]{\spacefactor3000\relax}%
\providecommand \BibitemShut  [1]{\csname bibitem#1\endcsname}%
\let\auto@bib@innerbib\@empty
\bibitem [{\citenamefont {Pauli}(1927)}]{nuclear1}%
  \BibitemOpen
  \bibfield  {author} {\bibinfo {author} {\bibfnamefont {W.}~\bibnamefont
  {Pauli}},\ }\bibfield  {title} {\bibinfo {title} {{Zur Quantenmechanik des
  magnetischen Elektrons}},\ }\href {https://doi.org/10.1007/bf01397326}
  {\bibfield  {journal} {\bibinfo  {journal} {Z. Phys.}\ }\textbf {\bibinfo
  {volume} {43}},\ \bibinfo {pages} {601} (\bibinfo {year} {1927})}\BibitemShut
  {NoStop}%
\bibitem [{\citenamefont {Rosen}\ and\ \citenamefont
  {Morse}(1932)}]{nuclear12}%
  \BibitemOpen
  \bibfield  {author} {\bibinfo {author} {\bibfnamefont {N.}~\bibnamefont
  {Rosen}}\ and\ \bibinfo {author} {\bibfnamefont {P.~M.}\ \bibnamefont
  {Morse}},\ }\bibfield  {title} {\bibinfo {title} {On the vibrations of
  polyatomic molecules},\ }\href@noop {} {\bibfield  {journal} {\bibinfo
  {journal} {Phys. Rev.}\ }\textbf {\bibinfo {volume} {42}},\ \bibinfo {pages}
  {210} (\bibinfo {year} {1932})}\BibitemShut {NoStop}%
\bibitem [{\citenamefont {Schiff}\ \emph {et~al.}(1940)\citenamefont {Schiff},
  \citenamefont {Snyder},\ and\ \citenamefont {Weinberg}}]{nuclear7}%
  \BibitemOpen
  \bibfield  {author} {\bibinfo {author} {\bibfnamefont {L.}~\bibnamefont
  {Schiff}}, \bibinfo {author} {\bibfnamefont {H.}~\bibnamefont {Snyder}},\
  and\ \bibinfo {author} {\bibfnamefont {J.}~\bibnamefont {Weinberg}},\
  }\bibfield  {title} {\bibinfo {title} {On the existence of stationary states
  of the mesotron field},\ }\href@noop {} {\bibfield  {journal} {\bibinfo
  {journal} {Phys. Rev.}\ }\textbf {\bibinfo {volume} {57}},\ \bibinfo {pages}
  {315} (\bibinfo {year} {1940})}\BibitemShut {NoStop}%
\bibitem [{\citenamefont {Rizov}\ \emph {et~al.}(1985)\citenamefont {Rizov},
  \citenamefont {Sazdjian},\ and\ \citenamefont {Todorov}}]{nuclear5}%
  \BibitemOpen
  \bibfield  {author} {\bibinfo {author} {\bibfnamefont {V.}~\bibnamefont
  {Rizov}}, \bibinfo {author} {\bibfnamefont {H.}~\bibnamefont {Sazdjian}},\
  and\ \bibinfo {author} {\bibfnamefont {I.~T.}\ \bibnamefont {Todorov}},\
  }\bibfield  {title} {\bibinfo {title} {On the relativistic quantum mechanics
  of two interacting spinless particles},\ }\href@noop {} {\bibfield  {journal}
  {\bibinfo  {journal} {Ann. Phys. (N.Y.)}\ }\textbf {\bibinfo {volume}
  {165}},\ \bibinfo {pages} {59} (\bibinfo {year} {1985})}\BibitemShut
  {NoStop}%
\bibitem [{\citenamefont {Sazdjian}(1986)}]{nuclear6}%
  \BibitemOpen
  \bibfield  {author} {\bibinfo {author} {\bibfnamefont {H.}~\bibnamefont
  {Sazdjian}},\ }\bibfield  {title} {\bibinfo {title} {Relativistic wave
  equations for the dynamics of two interacting particles},\ }\href@noop {}
  {\bibfield  {journal} {\bibinfo  {journal} {Phys. Rev. D}\ }\textbf {\bibinfo
  {volume} {33}},\ \bibinfo {pages} {3401} (\bibinfo {year}
  {1986})}\BibitemShut {NoStop}%
\bibitem [{\citenamefont {Sazdjian}(1988)}]{nuclear13}%
  \BibitemOpen
  \bibfield  {author} {\bibinfo {author} {\bibfnamefont {H.}~\bibnamefont
  {Sazdjian}},\ }\bibfield  {title} {\bibinfo {title} {The scalar product in
  two-particle relativistic quantum mechanics},\ }\href@noop {} {\bibfield
  {journal} {\bibinfo  {journal} {J. Math. Phys. (N.Y.)}\ }\textbf {\bibinfo
  {volume} {29}},\ \bibinfo {pages} {1620} (\bibinfo {year}
  {1988})}\BibitemShut {NoStop}%
\bibitem [{\citenamefont {Mourad}\ and\ \citenamefont
  {Sazdjian}(1994)}]{nuclear10}%
  \BibitemOpen
  \bibfield  {author} {\bibinfo {author} {\bibfnamefont {J.}~\bibnamefont
  {Mourad}}\ and\ \bibinfo {author} {\bibfnamefont {H.}~\bibnamefont
  {Sazdjian}},\ }\bibfield  {title} {\bibinfo {title} {{The two-fermion
  relativistic wave equations of constraint theory in the
  Pauli--Schr{\"o}dinger form}},\ }\href@noop {} {\bibfield  {journal}
  {\bibinfo  {journal} {J. Math. Phys. (N.Y.)}\ }\textbf {\bibinfo {volume}
  {35}},\ \bibinfo {pages} {6379} (\bibinfo {year} {1994})}\BibitemShut
  {NoStop}%
\bibitem [{\citenamefont {Formanek}\ \emph {et~al.}(2004)\citenamefont
  {Formanek}, \citenamefont {Lombard},\ and\ \citenamefont
  {Mare{\v{s}}}}]{nuclear11}%
  \BibitemOpen
  \bibfield  {author} {\bibinfo {author} {\bibfnamefont {J.}~\bibnamefont
  {Formanek}}, \bibinfo {author} {\bibfnamefont {R.}~\bibnamefont {Lombard}},\
  and\ \bibinfo {author} {\bibfnamefont {J.}~\bibnamefont {Mare{\v{s}}}},\
  }\bibfield  {title} {\bibinfo {title} {Wave equations with energy-dependent
  potentials},\ }\href@noop {} {\bibfield  {journal} {\bibinfo  {journal}
  {Czech. J. Phys.}\ }\textbf {\bibinfo {volume} {54}},\ \bibinfo {pages} {289}
  (\bibinfo {year} {2004})}\BibitemShut {NoStop}%
\bibitem [{\citenamefont {Garcia-Martinez}\ \emph {et~al.}(2009)\citenamefont
  {Garcia-Martinez}, \citenamefont {Garcia-Ravelo}, \citenamefont {Pena},\ and\
  \citenamefont {Schulze-Halberg}}]{nuclear3}%
  \BibitemOpen
  \bibfield  {author} {\bibinfo {author} {\bibfnamefont {J.}~\bibnamefont
  {Garcia-Martinez}}, \bibinfo {author} {\bibfnamefont {J.}~\bibnamefont
  {Garcia-Ravelo}}, \bibinfo {author} {\bibfnamefont {J.}~\bibnamefont
  {Pena}},\ and\ \bibinfo {author} {\bibfnamefont {A.}~\bibnamefont
  {Schulze-Halberg}},\ }\bibfield  {title} {\bibinfo {title} {Exactly solvable
  energy-dependent potentials},\ }\href@noop {} {\bibfield  {journal} {\bibinfo
   {journal} {Phys. Lett. A}\ }\textbf {\bibinfo {volume} {373}},\ \bibinfo
  {pages} {3619} (\bibinfo {year} {2009})}\BibitemShut {NoStop}%
\bibitem [{\citenamefont {Lombard}\ \emph {et~al.}(2007)\citenamefont
  {Lombard}, \citenamefont {Mare{\v{s}}},\ and\ \citenamefont
  {Volpe}}]{nuclear9}%
  \BibitemOpen
  \bibfield  {author} {\bibinfo {author} {\bibfnamefont {R.}~\bibnamefont
  {Lombard}}, \bibinfo {author} {\bibfnamefont {J.}~\bibnamefont
  {Mare{\v{s}}}},\ and\ \bibinfo {author} {\bibfnamefont {C.}~\bibnamefont
  {Volpe}},\ }\bibfield  {title} {\bibinfo {title} {Wave equation with
  energy-dependent potentials for confined systems},\ }\href@noop {} {\bibfield
   {journal} {\bibinfo  {journal} {J. Phys. G}\ }\textbf {\bibinfo {volume}
  {34}},\ \bibinfo {pages} {1879} (\bibinfo {year} {2007})}\BibitemShut
  {NoStop}%
\bibitem [{\citenamefont {Langueur}\ \emph {et~al.}(2019)\citenamefont
  {Langueur}, \citenamefont {Merad},\ and\ \citenamefont {Hamil}}]{nuclear4}%
  \BibitemOpen
  \bibfield  {author} {\bibinfo {author} {\bibfnamefont {O.}~\bibnamefont
  {Langueur}}, \bibinfo {author} {\bibfnamefont {M.}~\bibnamefont {Merad}},\
  and\ \bibinfo {author} {\bibfnamefont {B.}~\bibnamefont {Hamil}},\ }\bibfield
   {title} {\bibinfo {title} {Dkp equation with energy dependent potentials},\
  }\href@noop {} {\bibfield  {journal} {\bibinfo  {journal} {Commun. Theor.
  Phys.}\ }\textbf {\bibinfo {volume} {71}},\ \bibinfo {pages} {1069} (\bibinfo
  {year} {2019})}\BibitemShut {NoStop}%
\bibitem [{\citenamefont {Schulze-Halberg}\ and\ \citenamefont
  {Ye{\c{s}}ilta{\c{s}}}(2018)}]{nuclear75}%
  \BibitemOpen
  \bibfield  {author} {\bibinfo {author} {\bibfnamefont {A.}~\bibnamefont
  {Schulze-Halberg}}\ and\ \bibinfo {author} {\bibfnamefont
  {{\"O}.}~\bibnamefont {Ye{\c{s}}ilta{\c{s}}}},\ }\bibfield  {title} {\bibinfo
  {title} {{Generalized Schr{\"o}dinger equations with energy-dependent
  potentials: Formalism and applications}},\ }\href@noop {} {\bibfield
  {journal} {\bibinfo  {journal} {J. Math. Phys. (N.Y.)}\ }\textbf {\bibinfo
  {volume} {59}} (\bibinfo {year} {2018})}\BibitemShut {NoStop}%
\bibitem [{\citenamefont {Schulze-Halberg}(2020)}]{nuclear8}%
  \BibitemOpen
  \bibfield  {author} {\bibinfo {author} {\bibfnamefont {A.}~\bibnamefont
  {Schulze-Halberg}},\ }\bibfield  {title} {\bibinfo {title} {{Higher-order
  Darboux transformations and Wronskian representations for Schr{\"o}dinger
  equations with quadratically energy-dependent potentials}},\ }\href@noop {}
  {\bibfield  {journal} {\bibinfo  {journal} {J. Math. Phys. (N.Y.)}\ }\textbf
  {\bibinfo {volume} {61}} (\bibinfo {year} {2020})}\BibitemShut {NoStop}%
\bibitem [{\citenamefont {Borrego-Morell}\ \emph {et~al.}(2020)\citenamefont
  {Borrego-Morell}, \citenamefont {Bracciali},\ and\ \citenamefont
  {Sri~Ranga}}]{nuclear2}%
  \BibitemOpen
  \bibfield  {author} {\bibinfo {author} {\bibfnamefont {J.~A.}\ \bibnamefont
  {Borrego-Morell}}, \bibinfo {author} {\bibfnamefont {C.~F.}\ \bibnamefont
  {Bracciali}},\ and\ \bibinfo {author} {\bibfnamefont {A.}~\bibnamefont
  {Sri~Ranga}},\ }\bibfield  {title} {\bibinfo {title} {{On an energy-dependent
  quantum system with solutions in terms of a class of hypergeometric
  para-orthogonal polynomials on the unit circle}},\ }\href
  {https://www.mdpi.com/2227-7390/8/7/1161} {\bibfield  {journal} {\bibinfo
  {journal} {Mathematics}\ }\textbf {\bibinfo {volume} {8}} (\bibinfo {year}
  {2020})}\BibitemShut {NoStop}%
\bibitem [{\citenamefont {Kokkotas}\ and\ \citenamefont
  {Schmidt}(1999)}]{Kokkotas:1999bd}%
  \BibitemOpen
  \bibfield  {author} {\bibinfo {author} {\bibfnamefont {K.~D.}\ \bibnamefont
  {Kokkotas}}\ and\ \bibinfo {author} {\bibfnamefont {B.~G.}\ \bibnamefont
  {Schmidt}},\ }\bibfield  {title} {\bibinfo {title} {{Quasinormal modes of
  stars and black holes}},\ }\href@noop {} {\bibfield  {journal} {\bibinfo
  {journal} {Living Rev. Relativity}\ }\textbf {\bibinfo {volume} {2}},\
  \bibinfo {pages} {2} (\bibinfo {year} {1999})}\BibitemShut {NoStop}%
\bibitem [{\citenamefont {Nollert}(1999)}]{Nollert:1999ji}%
  \BibitemOpen
  \bibfield  {author} {\bibinfo {author} {\bibfnamefont {H.-P.}\ \bibnamefont
  {Nollert}},\ }\bibfield  {title} {\bibinfo {title} {{TOPICAL REVIEW:
  Quasinormal modes: The characteristic `sound' of black holes and neutron
  stars}},\ }\href {https://doi.org/10.1088/0264-9381/16/12/201} {\bibfield
  {journal} {\bibinfo  {journal} {Classical Quantum Gravity}\ }\textbf
  {\bibinfo {volume} {16}},\ \bibinfo {pages} {R159} (\bibinfo {year}
  {1999})}\BibitemShut {NoStop}%
\bibitem [{\citenamefont {Berti}\ \emph {et~al.}(2009)\citenamefont {Berti},
  \citenamefont {Cardoso},\ and\ \citenamefont {Starinets}}]{Berti:2009kk}%
  \BibitemOpen
  \bibfield  {author} {\bibinfo {author} {\bibfnamefont {E.}~\bibnamefont
  {Berti}}, \bibinfo {author} {\bibfnamefont {V.}~\bibnamefont {Cardoso}},\
  and\ \bibinfo {author} {\bibfnamefont {A.~O.}\ \bibnamefont {Starinets}},\
  }\bibfield  {title} {\bibinfo {title} {{Quasinormal modes of black holes and
  black branes}},\ }\href {https://doi.org/10.1088/0264-9381/26/16/163001}
  {\bibfield  {journal} {\bibinfo  {journal} {Classical Quantum Gravity}\
  }\textbf {\bibinfo {volume} {26}},\ \bibinfo {pages} {163001} (\bibinfo
  {year} {2009})},\ \Eprint {https://arxiv.org/abs/0905.2975} {arXiv:0905.2975
  [gr-qc]} \BibitemShut {NoStop}%
\bibitem [{\citenamefont {Konoplya}\ and\ \citenamefont
  {Zhidenko}(2011)}]{Konoplya:2011qq}%
  \BibitemOpen
  \bibfield  {author} {\bibinfo {author} {\bibfnamefont {R.~A.}\ \bibnamefont
  {Konoplya}}\ and\ \bibinfo {author} {\bibfnamefont {A.}~\bibnamefont
  {Zhidenko}},\ }\bibfield  {title} {\bibinfo {title} {{Quasinormal modes of
  black holes: From astrophysics to string theory}},\ }\href
  {https://doi.org/10.1103/RevModPhys.83.793} {\bibfield  {journal} {\bibinfo
  {journal} {Rev. Mod. Phys.}\ }\textbf {\bibinfo {volume} {83}},\ \bibinfo
  {pages} {793} (\bibinfo {year} {2011})},\ \Eprint
  {https://arxiv.org/abs/1102.4014} {arXiv:1102.4014 [gr-qc]} \BibitemShut
  {NoStop}%
\bibitem [{\citenamefont {Barcelo}\ \emph {et~al.}(2005)\citenamefont
  {Barcelo}, \citenamefont {Liberati},\ and\ \citenamefont
  {Visser}}]{Barcelo:2005fc}%
  \BibitemOpen
  \bibfield  {author} {\bibinfo {author} {\bibfnamefont {C.}~\bibnamefont
  {Barcelo}}, \bibinfo {author} {\bibfnamefont {S.}~\bibnamefont {Liberati}},\
  and\ \bibinfo {author} {\bibfnamefont {M.}~\bibnamefont {Visser}},\
  }\bibfield  {title} {\bibinfo {title} {{Analogue gravity}},\ }\href
  {https://doi.org/10.12942/lrr-2005-12} {\bibfield  {journal} {\bibinfo
  {journal} {Living Rev. Relativity}\ }\textbf {\bibinfo {volume} {8}},\
  \bibinfo {pages} {12} (\bibinfo {year} {2005})},\ \Eprint
  {https://arxiv.org/abs/gr-qc/0505065} {arXiv:gr-qc/0505065} \BibitemShut
  {NoStop}%
\bibitem [{\citenamefont
  {Sommerfeld}(1916)}]{https://doi.org/10.1002/andp.19163561702}%
  \BibitemOpen
  \bibfield  {author} {\bibinfo {author} {\bibfnamefont {A.}~\bibnamefont
  {Sommerfeld}},\ }\bibfield  {title} {\bibinfo {title} {Zur quantentheorie der
  spektrallinien},\ }\href
  {https://doi.org/https://doi.org/10.1002/andp.19163561702} {\bibfield
  {journal} {\bibinfo  {journal} {Ann. Phys. (Berlin)}\ }\textbf {\bibinfo
  {volume} {356}},\ \bibinfo {pages} {1} (\bibinfo {year} {1916})}\BibitemShut
  {NoStop}%
\bibitem [{\citenamefont {Gamow}(1928)}]{Gamow:1928zz}%
  \BibitemOpen
  \bibfield  {author} {\bibinfo {author} {\bibfnamefont {G.}~\bibnamefont
  {Gamow}},\ }\bibfield  {title} {\bibinfo {title} {{Zur Quantentheorie des
  Atomkernes}},\ }\href {https://doi.org/10.1007/BF01343196} {\bibfield
  {journal} {\bibinfo  {journal} {Z. Phys.}\ }\textbf {\bibinfo {volume}
  {51}},\ \bibinfo {pages} {204} (\bibinfo {year} {1928})}\BibitemShut
  {NoStop}%
\bibitem [{\citenamefont {Wheeler}(2015)}]{lieb2015studies}%
  \BibitemOpen
  \bibfield  {author} {\bibinfo {author} {\bibfnamefont {J.~A.}\ \bibnamefont
  {Wheeler}},\ }\href {https://press.princeton.edu/titles/861.html} {\emph
  {\bibinfo {title} {{Studies in Mathematical Physics: Essays in Honor of
  Valentine Bargmann}}}},\ Princeton Series in Physics\ (\bibinfo  {publisher}
  {Princeton University Press, Princeton, NJ,},\ \bibinfo {year} {2015})\ pp.\
  \bibinfo {pages} {351--422}\BibitemShut {NoStop}%
\bibitem [{\citenamefont {Chadan}\ and\ \citenamefont
  {Sabatier}(1989)}]{MR985100}%
  \BibitemOpen
  \bibfield  {author} {\bibinfo {author} {\bibfnamefont {K.}~\bibnamefont
  {Chadan}}\ and\ \bibinfo {author} {\bibfnamefont {P.~C.}\ \bibnamefont
  {Sabatier}},\ }\href {https://doi.org/10.1007/978-3-642-83317-5} {\emph
  {\bibinfo {title} {{Inverse Problems in Quantum Scattering Theory}}}},\
  \bibinfo {edition} {2nd}\ ed.,\ Texts and Monographs in Physics\ (\bibinfo
  {publisher} {Springer-Verlag},\ \bibinfo {address} {New York},\ \bibinfo
  {year} {1989})\BibitemShut {NoStop}%
\bibitem [{\citenamefont {{Lazenby}}\ and\ \citenamefont
  {{Griffiths}}(1980)}]{1980AmJPh..48..432L}%
  \BibitemOpen
  \bibfield  {author} {\bibinfo {author} {\bibfnamefont {J.~C.}\ \bibnamefont
  {{Lazenby}}}\ and\ \bibinfo {author} {\bibfnamefont {D.~J.}\ \bibnamefont
  {{Griffiths}}},\ }\bibfield  {title} {\bibinfo {title} {{Classical inverse
  scattering in one dimension}},\ }\href {https://doi.org/10.1119/1.11998}
  {\bibfield  {journal} {\bibinfo  {journal} {Am. J. Phys.}\ }\textbf {\bibinfo
  {volume} {48}},\ \bibinfo {pages} {432} (\bibinfo {year} {1980})}\BibitemShut
  {NoStop}%
\bibitem [{\citenamefont {{Gandhi}}\ and\ \citenamefont
  {{Efthimiou}}(2006)}]{2006AmJPh..74..638G}%
  \BibitemOpen
  \bibfield  {author} {\bibinfo {author} {\bibfnamefont {S.~C.}\ \bibnamefont
  {{Gandhi}}}\ and\ \bibinfo {author} {\bibfnamefont {C.~J.}\ \bibnamefont
  {{Efthimiou}}},\ }\bibfield  {title} {\bibinfo {title} {{Inversion of Gamow's
  formula and inverse scattering}},\ }\href {https://doi.org/10.1119/1.2190683}
  {\bibfield  {journal} {\bibinfo  {journal} {Am. J. Phys.}\ }\textbf {\bibinfo
  {volume} {74}},\ \bibinfo {pages} {638} (\bibinfo {year} {2006})},\ \Eprint
  {https://arxiv.org/abs/quant-ph/0503223} {quant-ph/0503223} \BibitemShut
  {NoStop}%
\bibitem [{\citenamefont {V\"olkel}\ and\ \citenamefont
  {Kokkotas}(2017)}]{Volkel:2017kfj}%
  \BibitemOpen
  \bibfield  {author} {\bibinfo {author} {\bibfnamefont {S.~H.}\ \bibnamefont
  {V\"olkel}}\ and\ \bibinfo {author} {\bibfnamefont {K.~D.}\ \bibnamefont
  {Kokkotas}},\ }\bibfield  {title} {\bibinfo {title} {{Ultra compact stars:
  Reconstructing the perturbation potential}},\ }\href
  {https://doi.org/10.1088/1361-6382/aa82de} {\bibfield  {journal} {\bibinfo
  {journal} {Classical Quantum Gravity}\ }\textbf {\bibinfo {volume} {34}},\
  \bibinfo {pages} {175015} (\bibinfo {year} {2017})},\ \Eprint
  {https://arxiv.org/abs/1704.07517} {arXiv:1704.07517 [gr-qc]} \BibitemShut
  {NoStop}%
\bibitem [{\citenamefont {V\"olkel}\ and\ \citenamefont
  {Kokkotas}(2018)}]{Volkel:2018hwb}%
  \BibitemOpen
  \bibfield  {author} {\bibinfo {author} {\bibfnamefont {S.~H.}\ \bibnamefont
  {V\"olkel}}\ and\ \bibinfo {author} {\bibfnamefont {K.~D.}\ \bibnamefont
  {Kokkotas}},\ }\bibfield  {title} {\bibinfo {title} {{Wormhole potentials and
  throats from quasi-normal modes}},\ }\href
  {https://doi.org/10.1088/1361-6382/aabce6} {\bibfield  {journal} {\bibinfo
  {journal} {Classical Quantum Gravity}\ }\textbf {\bibinfo {volume} {35}},\
  \bibinfo {pages} {105018} (\bibinfo {year} {2018})},\ \Eprint
  {https://arxiv.org/abs/1802.08525} {arXiv:1802.08525 [gr-qc]} \BibitemShut
  {NoStop}%
\bibitem [{\citenamefont {V\"olkel}\ and\ \citenamefont
  {Kokkotas}(2019)}]{Volkel:2019gpq}%
  \BibitemOpen
  \bibfield  {author} {\bibinfo {author} {\bibfnamefont {S.~H.}\ \bibnamefont
  {V\"olkel}}\ and\ \bibinfo {author} {\bibfnamefont {K.~D.}\ \bibnamefont
  {Kokkotas}},\ }\bibfield  {title} {\bibinfo {title} {{On the Inverse Spectrum
  Problem of Neutron Stars}},\ }\href
  {https://doi.org/10.1088/1361-6382/ab186e} {\bibfield  {journal} {\bibinfo
  {journal} {Classical Quantum Gravity}\ }\textbf {\bibinfo {volume} {36}},\
  \bibinfo {pages} {115002} (\bibinfo {year} {2019})},\ \Eprint
  {https://arxiv.org/abs/1901.11262} {arXiv:1901.11262 [gr-qc]} \BibitemShut
  {NoStop}%
\bibitem [{\citenamefont {Bonatsos}\ \emph {et~al.}(1992)\citenamefont
  {Bonatsos}, \citenamefont {Daskaloyannis},\ and\ \citenamefont
  {Kokkotas}}]{Bonatsos:1992qq}%
  \BibitemOpen
  \bibfield  {author} {\bibinfo {author} {\bibfnamefont {D.}~\bibnamefont
  {Bonatsos}}, \bibinfo {author} {\bibfnamefont {C.}~\bibnamefont
  {Daskaloyannis}},\ and\ \bibinfo {author} {\bibfnamefont {K.~D.}\
  \bibnamefont {Kokkotas}},\ }\bibfield  {title} {\bibinfo {title} {{WKB
  equivalent potentials for q deformed harmonic and anharmonic oscillators}},\
  }\href {https://doi.org/10.1063/1.529565} {\bibfield  {journal} {\bibinfo
  {journal} {J. Math. Phys. (N.Y.)}\ }\textbf {\bibinfo {volume} {33}},\
  \bibinfo {pages} {2958} (\bibinfo {year} {1992})}\BibitemShut {NoStop}%
\bibitem [{\citenamefont {Kokkotas}(1993)}]{Kokkotas:1993ef}%
  \BibitemOpen
  \bibfield  {author} {\bibinfo {author} {\bibfnamefont {K.~D.}\ \bibnamefont
  {Kokkotas}},\ }\bibfield  {title} {\bibinfo {title} {{Quasinormal modes of
  the Kerr-Newman black hole}},\ }\href {https://doi.org/10.1007/BF02822861}
  {\bibfield  {journal} {\bibinfo  {journal} {Nuovo Cimento B}\ }\textbf
  {\bibinfo {volume} {108}},\ \bibinfo {pages} {991} (\bibinfo {year}
  {1993})}\BibitemShut {NoStop}%
\bibitem [{\citenamefont {Konoplya}(2003)}]{Konoplya:2003ii}%
  \BibitemOpen
  \bibfield  {author} {\bibinfo {author} {\bibfnamefont {R.~A.}\ \bibnamefont
  {Konoplya}},\ }\bibfield  {title} {\bibinfo {title} {{Quasinormal behavior of
  the d-dimensional Schwarzschild black hole and higher order WKB approach}},\
  }\href {https://doi.org/10.1103/PhysRevD.68.024018} {\bibfield  {journal}
  {\bibinfo  {journal} {Phys. Rev. D}\ }\textbf {\bibinfo {volume} {68}},\
  \bibinfo {pages} {024018} (\bibinfo {year} {2003})},\ \Eprint
  {https://arxiv.org/abs/gr-qc/0303052} {arXiv:gr-qc/0303052} \BibitemShut
  {NoStop}%
\bibitem [{\citenamefont {Konoplya}\ \emph {et~al.}(2019)\citenamefont
  {Konoplya}, \citenamefont {Zhidenko},\ and\ \citenamefont
  {Zinhailo}}]{Konoplya:2019hlu}%
  \BibitemOpen
  \bibfield  {author} {\bibinfo {author} {\bibfnamefont {R.~A.}\ \bibnamefont
  {Konoplya}}, \bibinfo {author} {\bibfnamefont {A.}~\bibnamefont {Zhidenko}},\
  and\ \bibinfo {author} {\bibfnamefont {A.~F.}\ \bibnamefont {Zinhailo}},\
  }\bibfield  {title} {\bibinfo {title} {{Higher order WKB formula for
  quasinormal modes and grey-body factors: Recipes for quick and accurate
  calculations}},\ }\href {https://doi.org/10.1088/1361-6382/ab2e25} {\bibfield
   {journal} {\bibinfo  {journal} {Classical Quantum Gravity}\ }\textbf
  {\bibinfo {volume} {36}},\ \bibinfo {pages} {155002} (\bibinfo {year}
  {2019})},\ \Eprint {https://arxiv.org/abs/1904.10333} {arXiv:1904.10333
  [gr-qc]} \BibitemShut {NoStop}%
\bibitem [{\citenamefont {{P\"oschl, G. and Teller,
  E.}}(1933)}]{Poschl:1933zz}%
  \BibitemOpen
  \bibfield  {author} {\bibinfo {author} {\bibnamefont {{P\"oschl, G. and
  Teller, E.}}},\ }\bibfield  {title} {\bibinfo {title} {{Bemerkungen zur
  Quantenmechanik des anharmonischen Oszillators}},\ }\href
  {https://doi.org/10.1007/BF01331132} {\bibfield  {journal} {\bibinfo
  {journal} {Z. Phys.}\ }\textbf {\bibinfo {volume} {83}},\ \bibinfo {pages}
  {143} (\bibinfo {year} {1933})}\BibitemShut {NoStop}%
\bibitem [{\citenamefont {Bender}\ \emph {et~al.}(1999)\citenamefont {Bender},
  \citenamefont {Orszag},\ and\ \citenamefont {Orszag}}]{bender1999advanced}%
  \BibitemOpen
  \bibfield  {author} {\bibinfo {author} {\bibfnamefont {C.}~\bibnamefont
  {Bender}}, \bibinfo {author} {\bibfnamefont {S.}~\bibnamefont {Orszag}},\
  and\ \bibinfo {author} {\bibfnamefont {S.}~\bibnamefont {Orszag}},\ }\href
  {https://books.google.it/books?id=-yQXwhE6iWMC} {\emph {\bibinfo {title}
  {Advanced Mathematical Methods for Scientists and Engineers I: Asymptotic
  Methods and Perturbation Theory}}},\ Advanced Mathematical Methods for
  Scientists and Engineers\ (\bibinfo  {publisher} {Springer, New York},\
  \bibinfo {year} {1999})\BibitemShut {NoStop}%
\bibitem [{\citenamefont {{Karnakov}}\ and\ \citenamefont
  {{Krainov}}(2013)}]{2013waap.book.....K}%
  \BibitemOpen
  \bibfield  {author} {\bibinfo {author} {\bibfnamefont {B.~M.}\ \bibnamefont
  {{Karnakov}}}\ and\ \bibinfo {author} {\bibfnamefont {V.~P.}\ \bibnamefont
  {{Krainov}}},\ }\href {https://doi.org/10.1007/978-3-642-31558-9} {\emph
  {\bibinfo {title} {WKB Approximation in Atomic Physics: , ISBN
  978-3-642-31557-2.~Springer-Verlag Berlin Heidelberg, 2013}}}\ (\bibinfo
  {publisher} {Springer-Verlag Berlin Heidelberg},\ \bibinfo {year}
  {2013})\BibitemShut {NoStop}%
\bibitem [{\citenamefont {Albuquerque}\ \emph {et~al.}(2023)\citenamefont
  {Albuquerque}, \citenamefont {V\"olkel}, \citenamefont {Kokkotas},\ and\
  \citenamefont {Bezerra}}]{Albuquerque:2023lzw}%
  \BibitemOpen
  \bibfield  {author} {\bibinfo {author} {\bibfnamefont {S.}~\bibnamefont
  {Albuquerque}}, \bibinfo {author} {\bibfnamefont {S.~H.}\ \bibnamefont
  {V\"olkel}}, \bibinfo {author} {\bibfnamefont {K.~D.}\ \bibnamefont
  {Kokkotas}},\ and\ \bibinfo {author} {\bibfnamefont {V.~B.}\ \bibnamefont
  {Bezerra}},\ }\bibfield  {title} {\bibinfo {title} {{Inverse problem of
  analog gravity systems}},\ }\href
  {https://doi.org/10.1103/PhysRevD.108.124053} {\bibfield  {journal} {\bibinfo
   {journal} {Phys. Rev. D}\ }\textbf {\bibinfo {volume} {108}},\ \bibinfo
  {pages} {124053} (\bibinfo {year} {2023})},\ \Eprint
  {https://arxiv.org/abs/2309.11168} {arXiv:2309.11168 [gr-qc]} \BibitemShut
  {NoStop}%
\bibitem [{\citenamefont {V\"olkel}\ \emph {et~al.}(2019)\citenamefont
  {V\"olkel}, \citenamefont {Konoplya},\ and\ \citenamefont
  {Kokkotas}}]{Volkel:2019ahb}%
  \BibitemOpen
  \bibfield  {author} {\bibinfo {author} {\bibfnamefont {S.~H.}\ \bibnamefont
  {V\"olkel}}, \bibinfo {author} {\bibfnamefont {R.}~\bibnamefont {Konoplya}},\
  and\ \bibinfo {author} {\bibfnamefont {K.~D.}\ \bibnamefont {Kokkotas}},\
  }\bibfield  {title} {\bibinfo {title} {{Inverse problem for Hawking
  radiation}},\ }\href {https://doi.org/10.1103/PhysRevD.99.104025} {\bibfield
  {journal} {\bibinfo  {journal} {Phys. Rev. D}\ }\textbf {\bibinfo {volume}
  {99}},\ \bibinfo {pages} {104025} (\bibinfo {year} {2019})},\ \Eprint
  {https://arxiv.org/abs/1902.07611} {arXiv:1902.07611 [gr-qc]} \BibitemShut
  {NoStop}%
\bibitem [{\citenamefont {Kokkotas}(1991)}]{Kokkotas:1991vz}%
  \BibitemOpen
  \bibfield  {author} {\bibinfo {author} {\bibfnamefont {K.~D.}\ \bibnamefont
  {Kokkotas}},\ }\bibfield  {title} {\bibinfo {title} {{Normal modes of the
  Kerr black hole}},\ }\href {https://doi.org/10.1088/0264-9381/8/12/006}
  {\bibfield  {journal} {\bibinfo  {journal} {Classical Quantum Gravity}\
  }\textbf {\bibinfo {volume} {8}},\ \bibinfo {pages} {2217} (\bibinfo {year}
  {1991})}\BibitemShut {NoStop}%
\bibitem [{\citenamefont {Formánek}\ \emph {et~al.}(2004)\citenamefont
  {Formánek}, \citenamefont {Lombard},\ and\ \citenamefont
  {Mareš}}]{Form_nek_2004}%
  \BibitemOpen
  \bibfield  {author} {\bibinfo {author} {\bibfnamefont {J.}~\bibnamefont
  {Formánek}}, \bibinfo {author} {\bibfnamefont {R.}~\bibnamefont {Lombard}},\
  and\ \bibinfo {author} {\bibfnamefont {J.}~\bibnamefont {Mareš}},\
  }\bibfield  {title} {\bibinfo {title} {{Wave Equations with Energy-Dependent
  Potentials}},\ }\href {https://doi.org/10.1023/b:cjop.0000018127.95600.a3}
  {\bibfield  {journal} {\bibinfo  {journal} {Czech. J. Phys.}\ }\textbf
  {\bibinfo {volume} {54}},\ \bibinfo {pages} {289–316} (\bibinfo {year}
  {2004})}\BibitemShut {NoStop}%
\bibitem [{\citenamefont {Schutz}\ and\ \citenamefont
  {Will}(1985)}]{Schutz:1985km}%
  \BibitemOpen
  \bibfield  {author} {\bibinfo {author} {\bibfnamefont {B.~F.}\ \bibnamefont
  {Schutz}}\ and\ \bibinfo {author} {\bibfnamefont {C.~M.}\ \bibnamefont
  {Will}},\ }\bibfield  {title} {\bibinfo {title} {{Black hole normal modes: A
  semianalytic approach}},\ }\href {https://doi.org/10.1086/184453} {\bibfield
  {journal} {\bibinfo  {journal} {Astrophys. J. Lett.}\ }\textbf {\bibinfo
  {volume} {291}},\ \bibinfo {pages} {L33} (\bibinfo {year}
  {1985})}\BibitemShut {NoStop}%
\bibitem [{\citenamefont {Dong}\ \emph {et~al.}(2008)\citenamefont {Dong},
  \citenamefont {Qiang},\ and\ \citenamefont {García~Ravelo}}]{ptnuclear1}%
  \BibitemOpen
  \bibfield  {author} {\bibinfo {author} {\bibfnamefont {S.}~\bibnamefont
  {Dong}}, \bibinfo {author} {\bibfnamefont {W.-C.}\ \bibnamefont {Qiang}},\
  and\ \bibinfo {author} {\bibfnamefont {J.}~\bibnamefont {García~Ravelo}},\
  }\bibfield  {title} {\bibinfo {title} {{Analytical approximations to the
  Schrodinger equation for a second Poschl-Teller-like potential with
  centrifugal term}},\ }\href {https://doi.org/10.1142/S0217751X0803944X}
  {\bibfield  {journal} {\bibinfo  {journal} {Int. J. Mod. Phys. A}\ }\textbf
  {\bibinfo {volume} {23}},\ \bibinfo {pages} {1537} (\bibinfo {year}
  {2008})}\BibitemShut {NoStop}%
\bibitem [{\citenamefont {Dong}\ and\ \citenamefont
  {García~Ravelo}(2009)}]{ptnuclear2}%
  \BibitemOpen
  \bibfield  {author} {\bibinfo {author} {\bibfnamefont {S.}~\bibnamefont
  {Dong}}\ and\ \bibinfo {author} {\bibfnamefont {J.}~\bibnamefont
  {García~Ravelo}},\ }\bibfield  {title} {\bibinfo {title} {{Exact solutions
  of the Schrodinger equation with the Pöschl-Teller like potential}},\ }\href
  {https://doi.org/10.1142/S0217984909018047} {\bibfield  {journal} {\bibinfo
  {journal} {Mod. Phys. Lett. B}\ }\textbf {\bibinfo {volume} {23}},\ \bibinfo
  {pages} {603} (\bibinfo {year} {2009})}\BibitemShut {NoStop}%
\bibitem [{\citenamefont {Mashhoon}(1982)}]{Mashhoon:1982im}%
  \BibitemOpen
  \bibfield  {author} {\bibinfo {author} {\bibfnamefont {B.}~\bibnamefont
  {Mashhoon}},\ }\bibfield  {title} {\bibinfo {title} {Quasinormal modes of a
  black hole},\ }in\ \href@noop {} {\emph {\bibinfo {booktitle} {{3rd Marcel
  Grossmann Meeting on the Recent Developments of General Relativity}}}}\
  (\bibinfo {year} {1982})\BibitemShut {NoStop}%
\bibitem [{\citenamefont {Ferrari}\ and\ \citenamefont
  {Mashhoon}(1984)}]{Ferrari:1984ozr}%
  \BibitemOpen
  \bibfield  {author} {\bibinfo {author} {\bibfnamefont {V.}~\bibnamefont
  {Ferrari}}\ and\ \bibinfo {author} {\bibfnamefont {B.}~\bibnamefont
  {Mashhoon}},\ }\bibfield  {title} {\bibinfo {title} {{Oscillations of a black
  hole}},\ }\href {https://doi.org/10.1103/PhysRevLett.52.1361} {\bibfield
  {journal} {\bibinfo  {journal} {Phys. Rev. Lett.}\ }\textbf {\bibinfo
  {volume} {52}},\ \bibinfo {pages} {1361} (\bibinfo {year}
  {1984})}\BibitemShut {NoStop}%
\bibitem [{\citenamefont {Price}\ and\ \citenamefont
  {Khanna}(2017)}]{Price:2017cjr}%
  \BibitemOpen
  \bibfield  {author} {\bibinfo {author} {\bibfnamefont {R.~H.}\ \bibnamefont
  {Price}}\ and\ \bibinfo {author} {\bibfnamefont {G.}~\bibnamefont {Khanna}},\
  }\bibfield  {title} {\bibinfo {title} {{Gravitational wave sources:
  Reflections and echoes}},\ }\href {https://doi.org/10.1088/1361-6382/aa8f29}
  {\bibfield  {journal} {\bibinfo  {journal} {Classical Quantum Gravity}\
  }\textbf {\bibinfo {volume} {34}},\ \bibinfo {pages} {225005} (\bibinfo
  {year} {2017})},\ \Eprint {https://arxiv.org/abs/1702.04833}
  {arXiv:1702.04833 [gr-qc]} \BibitemShut {NoStop}%
\bibitem [{\citenamefont {V\"olkel}(2018)}]{Volkel:2018czg}%
  \BibitemOpen
  \bibfield  {author} {\bibinfo {author} {\bibfnamefont {S.~H.}\ \bibnamefont
  {V\"olkel}},\ }\bibfield  {title} {\bibinfo {title} {{Inverse spectrum
  problem for quasi-stationary states}},\ }\href
  {https://doi.org/10.1088/2399-6528/aaaee2} {\bibfield  {journal} {\bibinfo
  {journal} {J. Phys. Commun.}\ }\textbf {\bibinfo {volume} {2}},\ \bibinfo
  {pages} {025029} (\bibinfo {year} {2018})},\ \Eprint
  {https://arxiv.org/abs/1802.08684} {arXiv:1802.08684 [quant-ph]} \BibitemShut
  {NoStop}%
\end{thebibliography}%

\appendix

\section{SUPPLEMENTARY RESULTS}\label{appendix}

Throughout this paper, we divided our study into two main cases, the energy-dependent quadratic potential and the energy-dependent P\"oschl-Teller potential, with two scenarios of applications for each one; the effective inverse potential wells and the effective inverse potential barriers. In this appendix, we present complementary examples for each scenario.

\subsection{Quadratic potential}\label{app1}

\begin{figure}
\includegraphics[width=1.0\linewidth]{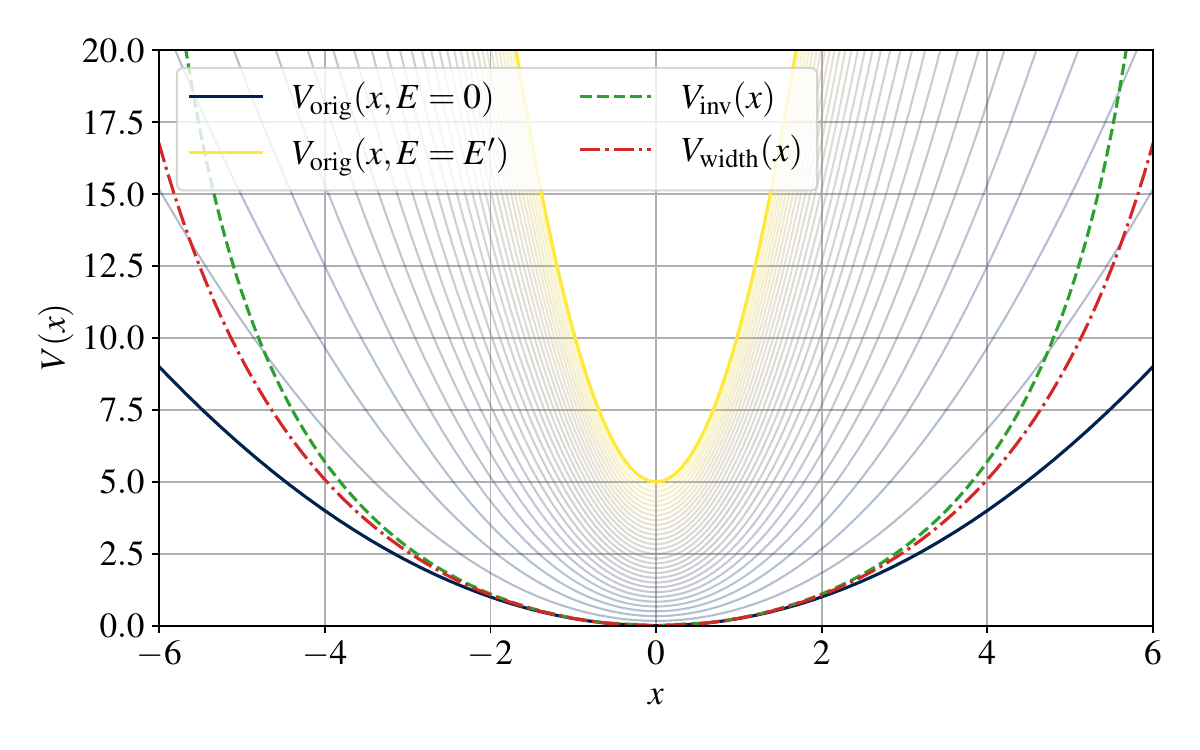}
\includegraphics[width=1.0\linewidth]{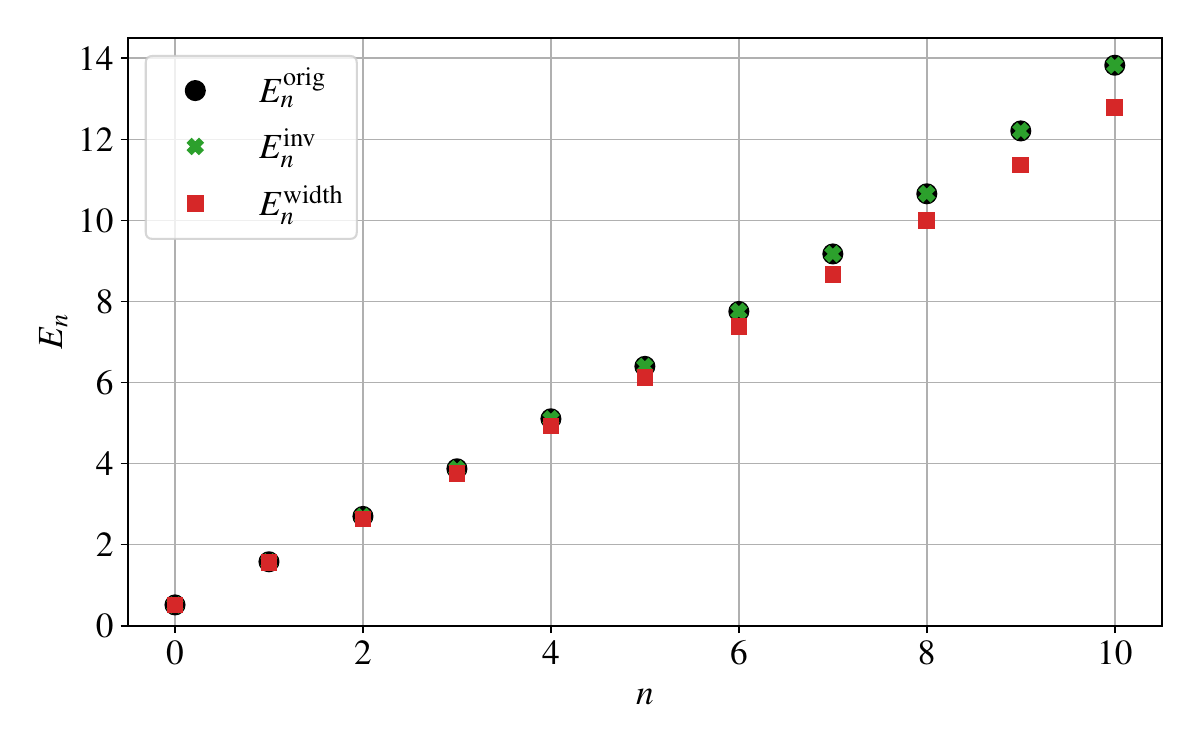}
\caption{Application of the inverse method to the energy-dependent quadratic potential Eq.~\eqref{parabollicdependent} for case $a/b>0$, with parameters $a=0.25$, $b=0.0125$ and $c=0.0125$. Top panel: The series of curves of varying colors indicate the chosen value of $E$ when computing the energy-dependent potential $V_\text{orig}(x, E)=V_\text{HO}(x,E)$ starting from $E=0$ to $E=E^\prime= 20 a/b$. The inverse potential (green dashed) is labeled as $V_\text{inv}(x)$ and the width-equivalent potential (red dotted dashed) is labeled as $V_\text{width}(x)$. Bottom panel: Here we show the spectrum of bound states for the original potential $E^\text{orig}_n$ (black circles), the inverse potential $E^\text{inv}_n$ (green cross) and width-equivalent potential $E^\text{width}_n$ (red squares). \label{parabollicresults2}}
\end{figure}

Complementing the discussion made in Fig.~\ref{parabollicresults}, we present here an extra example of an energy-dependent quadratic well. This second scenario is defined by $a/b>0$, for which the results and more details can be found in Fig.~\ref{parabollicresults2}. As can be seen in the top panel, the potential curves get increasingly more closed with higher values of $E$, until at some point, the turning points become asymptotically fixed at a certain distance, $x_{\rm limit}=\pm \sqrt{(1-c)/b}$. For these energies, the system behaves qualitatively like a particle in an infinite square well, which can also be seen from the bound states in the bottom panel. 

As in the previous case, we use the bound states as input for the inverse Bohr-Sommerfeld rule~\eqref{Excursion} to construct $V_\text{inv}(x)$. The width-equivalent potential $V_\text{width}(x)$ is included for comparison. 

In the bottom panel of Fig.~\ref{parabollicresults2}, we show the bound states for the three potentials. As in all previous cases, the bound states of the WKB-equivalent and original potentials agree very well. The asymptotic behavior of the overtones describes the transition from the spectrum of a quadratic potential to the one of a particle in an infinite square well ($E_{n}\propto n^2$).

\begin{figure}
\centering
\includegraphics[width=1.0\linewidth]{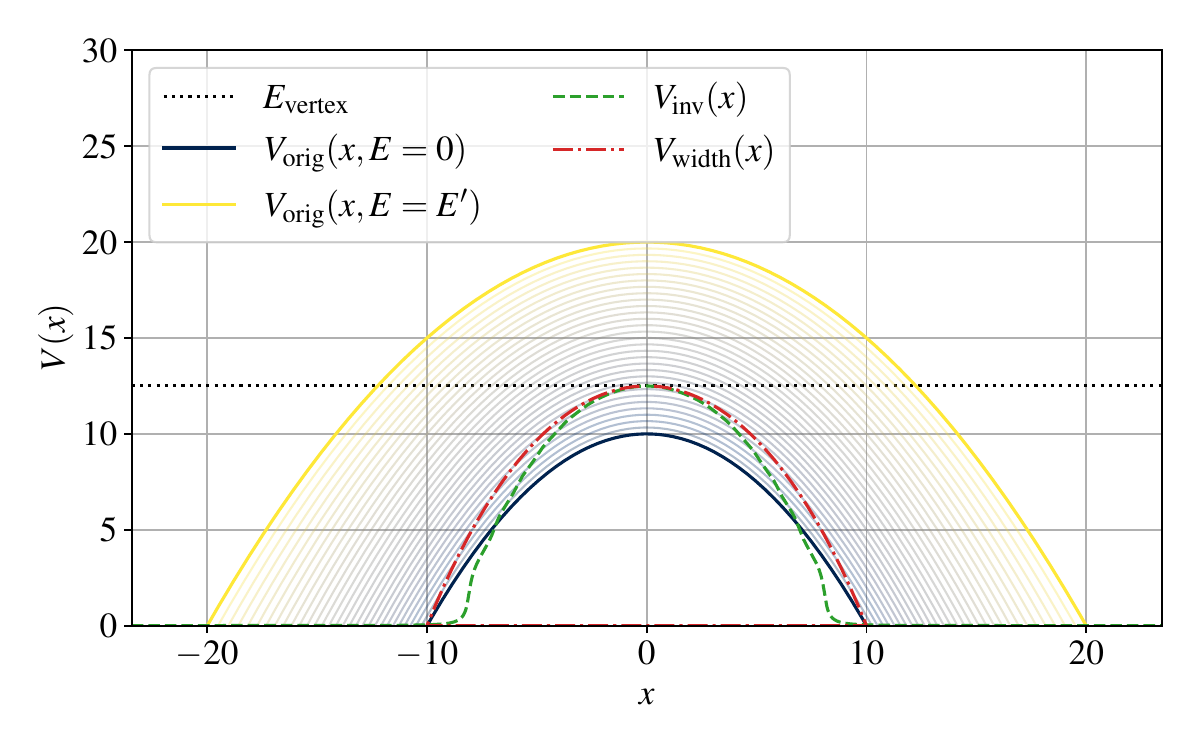}
\includegraphics[width=1.0\linewidth]{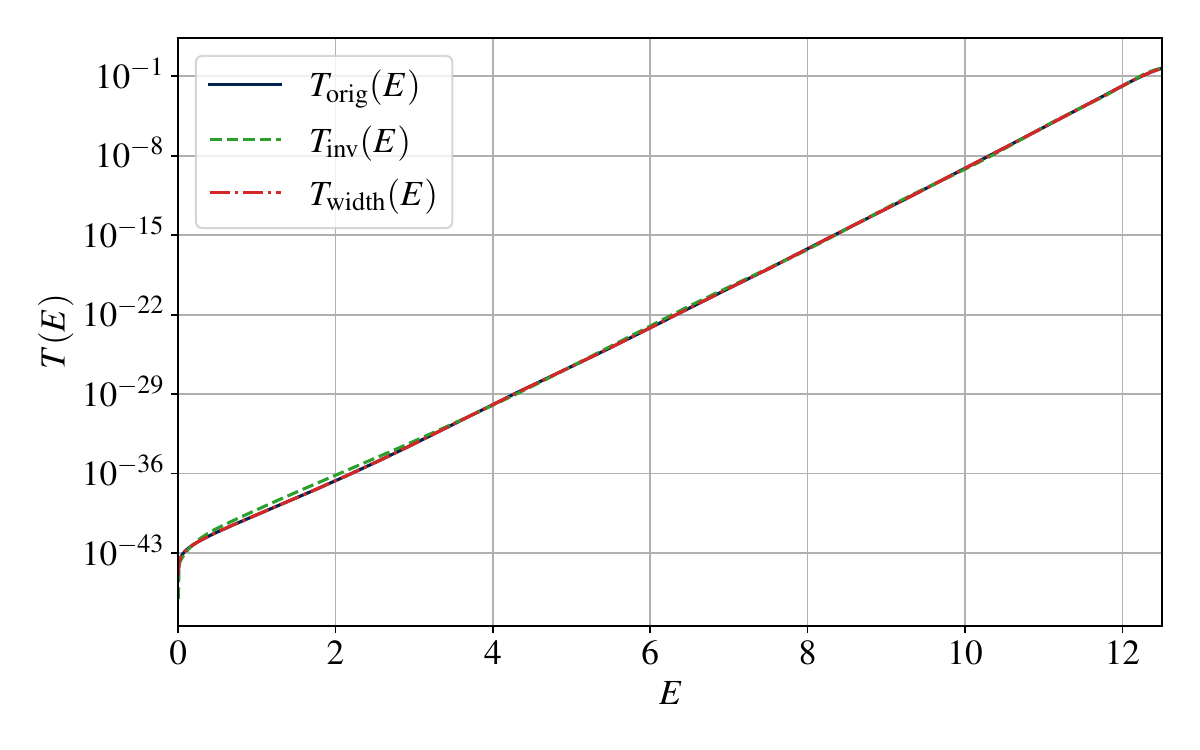}
\caption{Application of the inverse method to the energy-dependent quadratic potential barrier Eq.~\eqref{parabollicdependent} for case $a/b<0$, with parameters $a = -0.1$, $b = 0.001$, $c=0.2$, and $V_{0}=10$. Top panel: The series of curves of varying colors indicate the chosen value of $E$ when computing the energy-dependent potential $V_\text{orig}(x, E)=\bar{V}_\text{HO}(x,E)$ starting from $E=0$ to $E=E^\prime= 4 E_{\text{vertex}}$. The inverse potential (green dashed) is labeled as $V_\text{inv}(x)$ and the width equivalent potential (red dotted dashed) is labeled as $V_\text{width}(x)$. Bottom panel: Here we show the transmission for the original potential $T_{\rm orig}(E)$ (black solid line), the inverse potential  $T_{\rm inv}(E)$ (green dashed line) and the width-equivalent potential  $T_{\rm width}(E)$ (red dot-dashed lines).\label{inversemethodresults_V1}  \label{parabollicresults_T1}}
\end{figure}

Finally, we complement the discussion made in Fig.~\ref{inversemethodresults_V1} for energy-dependent quadratic barriers by considering $a/b<0$. The potential barrier curves get increasingly more open with increasing energy, which can be seen in the top panel of Fig.~\ref{inversemethodresults_V1}. The WKB-equivalent and width-equivalent potentials are also presented. As in the previous case in Fig.~\ref{parabollicresults_T2}, the difference between $V_\text{inv}(x)$ and the width equivalent potential $V_\text{width}(x)$ is apparent, although they become similar close to $E_{\rm vertex}$.  In the bottom panel of the same figure, we show the transmission associated with each potential. In this case, they are all very similar to each other.

\subsection{P\"oschl-Teller potential}\label{app2}

\begin{figure}
\centering
\includegraphics[width=1\linewidth]{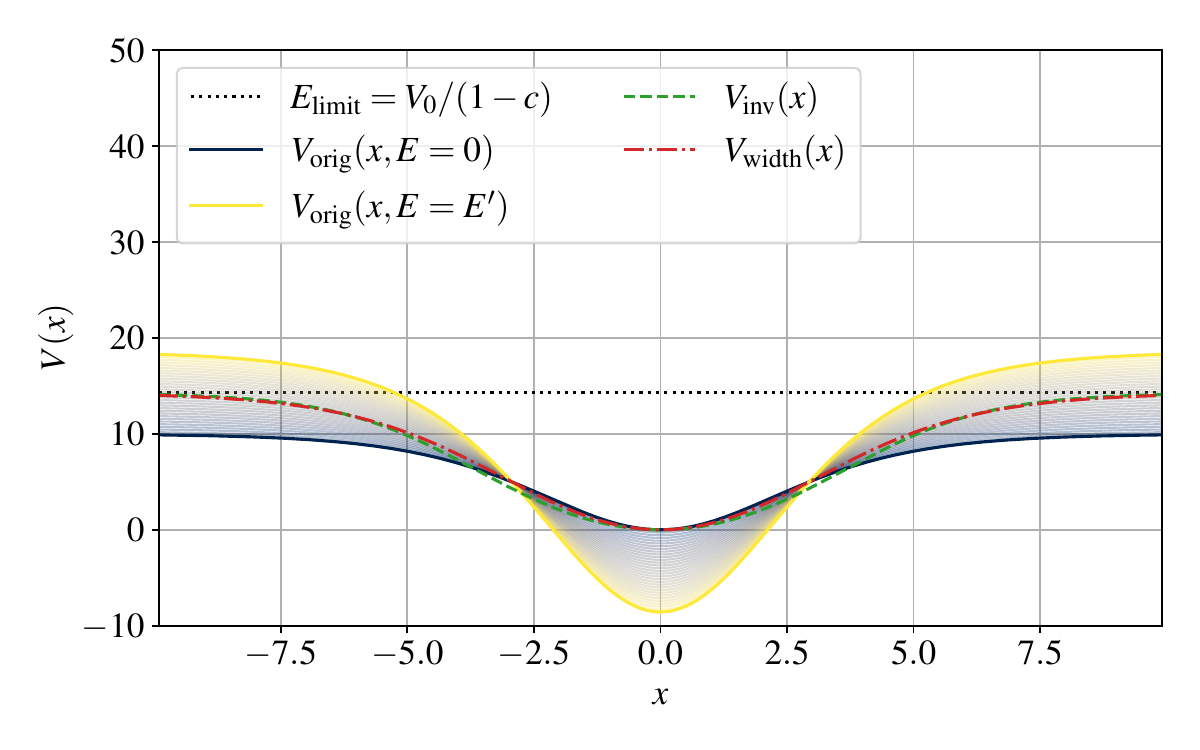}
\includegraphics[width=1\linewidth]{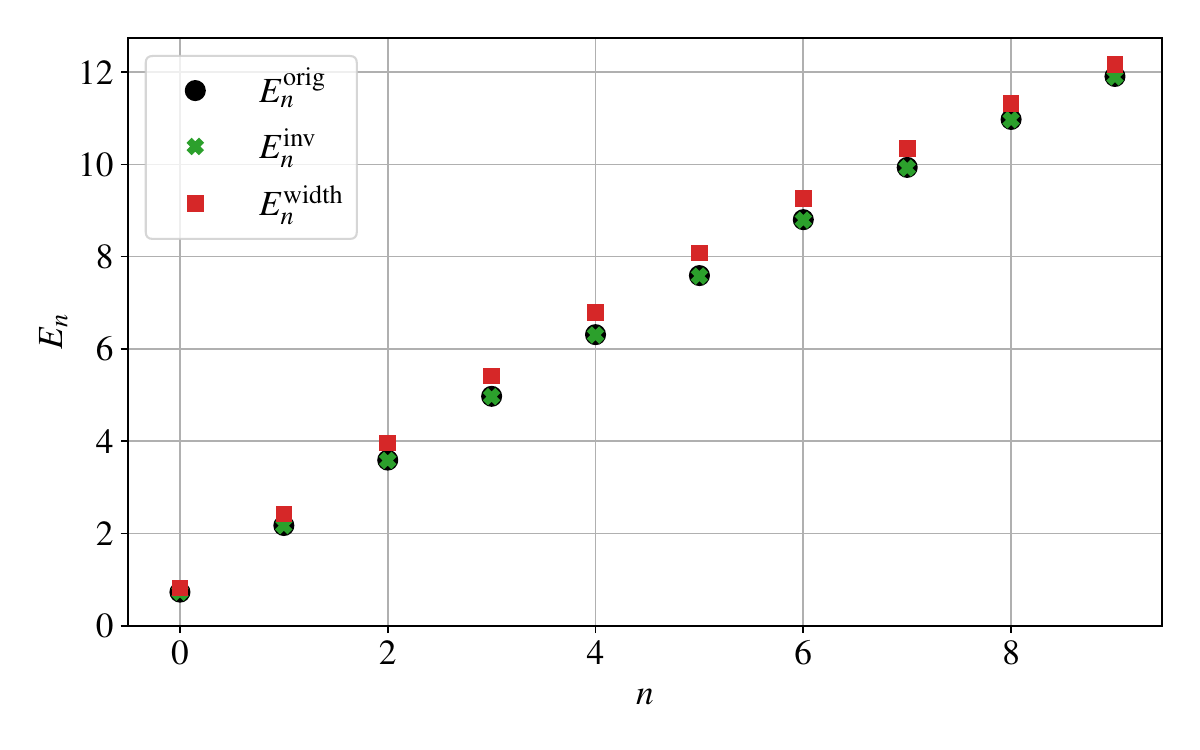}
\caption{Application of the inverse method to the energy-dependent P\"oschl-Teller potential Eq.~\eqref{ptdefinition} for case $a/b>0$, with parameters $a = -10$, $b=-0.6$, $c=0.3$, $k=0.3$, and $V_{0}=10$. Top panel: The series of curves of varying colors indicates the chosen value of $E$ when computing the energy-dependent potential $V_\text{orig}(x, E)=V_\text{PT}(x,E)$ starting from $E=0$ to $E=E^\prime= 2E_{\text{limit}}$. The inverse potential (green dashed) is labeled as $V_\text{inv}(x)$ and the width equivalent potential (red dotted dashed) is labeled as $V_\text{width}(x)$. Bottom panel: Here we show the spectrum of bound states for the original potential $E^\text{orig}_n$ (black circles), the inverse potential $E^\text{inv}_n$ (green cross) and width-equivalent potential $E^\text{width}_n$ (red squares).  \label{inversemethodresults_PT2}  \label{ptresults_PT2}}
\end{figure}

\begin{figure}[H]
\centering
\includegraphics[width=1\linewidth]{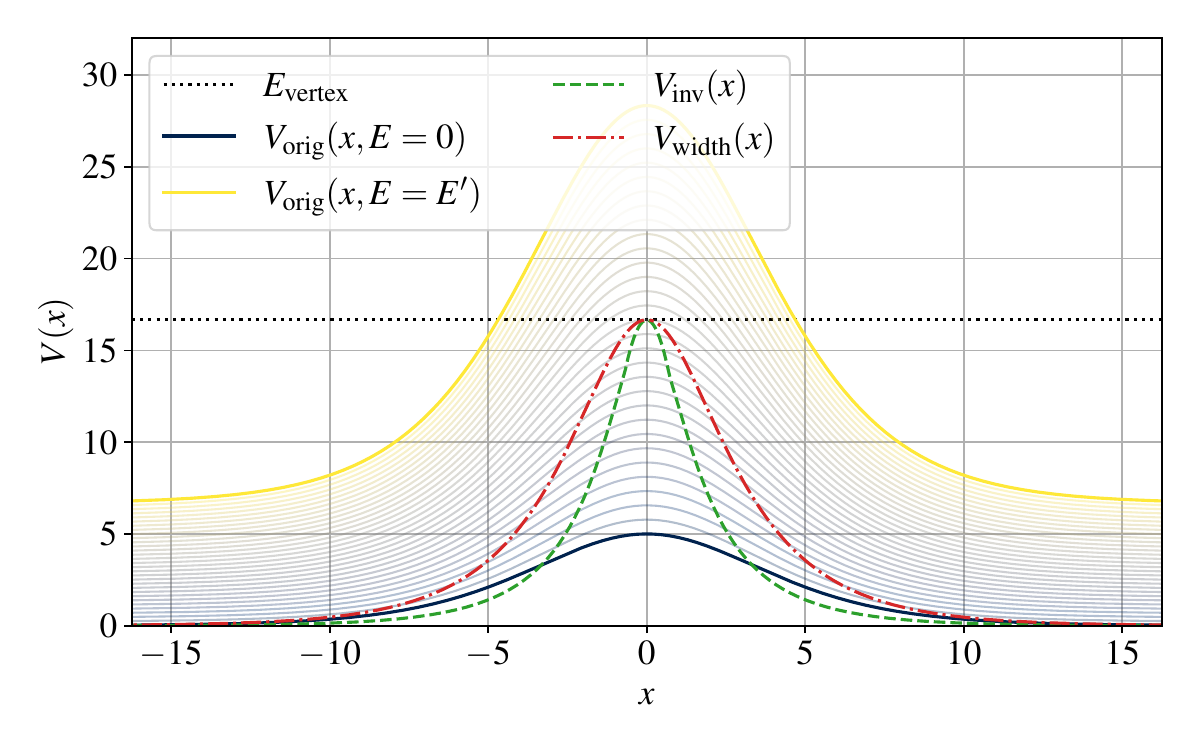}
\includegraphics[width=1\linewidth]{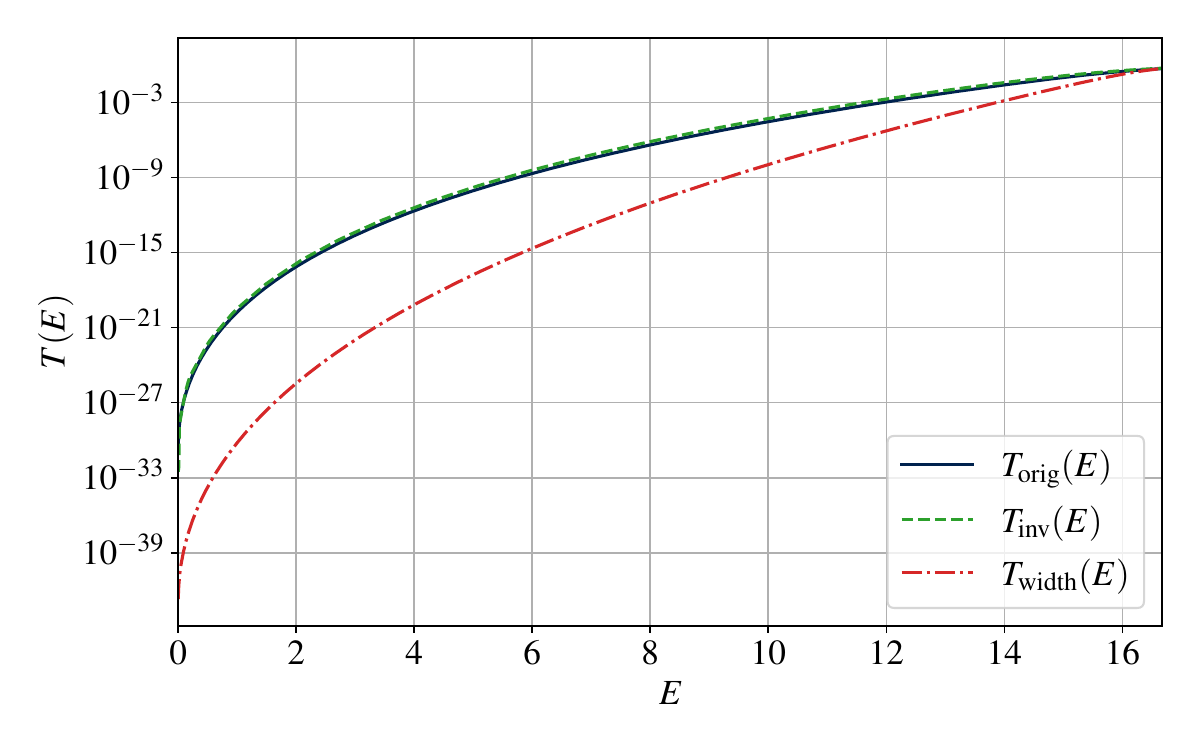}
\caption{Application of the inverse method to the energy-dependent P\"oschl-Teller potential Eq.~\eqref{ptdefinition} for case $a/b>0$, with parameters $a=5$, $b=0.5$, $c=0.2$, $k=0.2$, and $V_{0}=0$. Top panel: The series of curves of varying colors indicate the chosen value of $E$ when computing the energy-dependent potential $V_\text{orig}(x, E)=V_\text{PT}(x,E)$ starting from $E=0$ to $E=E^\prime = 2 E_{\text{vertex}}$. The inverse potential (green dashed) is labeled as $V_\text{inv}(x)$ and the width equivalent potential (red dotted dashed) is labeled as $V_\text{width}(x)$. Bottom panel: Bottom panel: Here we show the transmission for the original potential $T_{\rm orig}(E)$ (black solid line), the inverse potential  $T_{\rm inv}(E)$ (green dashed line) and the width-equivalent potential  $T_{\rm width}(E)$ (red dot-dashed lines). \label{ptbarrier2} }
\end{figure}

In Fig.~\ref{ptresults_PT2} and Fig.~\ref{ptbarrier2} we complement the previously introduced examples for energy-dependent P\"oschl-Teller potentials describing a potential-well reconstruction, and a potential-barrier reconstruction, respectively. 

\end{document}